\documentclass[aoas, preprint]{imsart}\usepackage[]{graphicx}\usepackage[]{color}
\makeatletter
\def\maxwidth{ %
  \ifdim\Gin@nat@width>\linewidth
    \linewidth
  \else
    \Gin@nat@width
  \fi
}
\makeatother

\definecolor{fgcolor}{rgb}{0.345, 0.345, 0.345}

\usepackage{framed}
\makeatletter
 {\par\unskip\endMakeFramed%
 \at@end@of@kframe}
\makeatother

\definecolor{shadecolor}{rgb}{.97, .97, .97}
\definecolor{messagecolor}{rgb}{0, 0, 0}
\definecolor{warningcolor}{rgb}{1, 0, 1}
\definecolor{errorcolor}{rgb}{1, 0, 0}
\newenvironment{knitrout}{}{} 

\usepackage{alltt}

\usepackage{amsthm,amsmath,natbib}
\RequirePackage[colorlinks,citecolor=blue,urlcolor=blue]{hyperref}

\arxiv{arXiv:1601.05788}

\startlocaldefs
\usepackage{enumitem}
\usepackage[hang, small,labelfont=bf,up,textfont=it,up]{caption} 
\usepackage{subcaption}
\usepackage{lineno}
\modulolinenumbers[5]

\usepackage{xr}
\endlocaldefs
\IfFileExists{upquote.sty}{\usepackage{upquote}}{}
\begin{document}

\begin{frontmatter}

\title{Automatic Matching of Bullet Land Impressions}\thanksref{T1}
\thankstext{T1}{Resources for this manuscript, including R code to reproduce all of the models, tables and figures, can be found at  \url{https://github.com/erichare/imaging-paper}.}
\runtitle{Automatic Matching of Bullet Land Impressions}
\runauthor{E. Hare et al.}

\begin{aug}
\author{\fnms{Eric} \snm{Hare}\corref{}\ead[label=e1]{erichare@iastate.edu}\thanksref{t1, m1}},
\author{\fnms{Heike} \snm{Hofmann}\ead[label=e2]{hofmann@iastate.edu}\thanksref{m1}}
\and
\author{\fnms{Alicia} \snm{Carriquiry}\ead[label=e3]{alicia@iastate.edu}\thanksref{m1}}

\thankstext{t1}{This research was partially funded by the Center for Statistics and Applications in Forensic Evidence (CSAFE) through Cooperative Agreement \#70NANB15H176 between NIST and Iowa State University, which includes activities carried out at Carnegie Mellon University, University of California Irvine, and University of Virginia.}
\affiliation{Iowa State University\thanksmark{m1}}

\address{Address of the Authors\\
Department of Statistics and Statistical Laboratory\\ Snedecor Hall\\
Ames Iowa 50011-1210\\
\printead{e1}\\
\phantom{E-mail:\ }\printead*{e2}\\
\phantom{E-mail:\ }\printead*{e3}}
\end{aug}
\begin{abstract}
In 2009, the National Academy of Sciences published a report questioning the scientific validity of many forensic methods including firearm examination. Firearm examination is a forensic tool used to help the court determine whether two bullets were fired from the same gun barrel. During the firing process, rifling, manufacturing defects, and impurities in the barrel create striation marks on the bullet. Identifying these striation markings in an attempt to match two bullets is one of the primary goals of firearm examination. We propose an automated framework for the analysis of the 3D surface measurements of bullet land impressions which transcribes the individual characteristics into a set of features that quantify their similarities. This makes identification of matches easier and allows for a quantification of both matches and matchability of barrels. The automatic matching routine we propose manages to (a) correctly identify land impressions (the surface between two bullet groove impressions) with too much damage to be suitable for comparison, and (b) correctly identify all 10,384 land-to-land matches of the James Hamby study \citep{hamby:2009}.
\end{abstract}



\end{frontmatter}


\section{Introduction}
Firearm examination is a forensic tool used to help the court determine whether two bullets were fired from the same gun barrel. This process has broad applicability in terms of convictions in the United States criminal justice system. Firearms identification has long been considered an accepted and reliable procedure, but in the past ten years has undergone more significant scrutiny. In 2005, in \emph{United States vs. Green}, the court ruled that the forensic expert could not confirm that the bullet casings came from a specific weapon with certainty, but could merely ``describe" other casings which are similar. Further court cases in the late 2000s expressed caution about the use of firearms identification evidence~\citep{giannelli:2011}.

In 2009, the National Academy of Sciences published a report~\citep{NAS:2009} questioning the scientific validity of many forensic methods including firearm examination. The report states that ``[m]uch forensic evidence -- including, for example, bite marks and firearm and toolmark identification -- is introduced in criminal trials without any meaningful scientific validation, determination of error rates , or reliability testing to explain the limits of the discipline."

Rifling, manufacturing defects, and impurities in a barrel create striation marks on the bullet during the firing process. These marks are assumed to be unique to the barrel, as described in a 1992 AFTE article~\citep{afte:1992}. ``The theory of identification as it pertains to the comparison of toolmarks enables opinions of common origin to be made when the \emph{unique surface contours} of two toolmarks are in sufficient agreement". The article goes on to state that ``Significance is determined by the comparative examination of two or more sets of surface contour patterns comprised of individual peaks, ridges and furrows."

From a statistical standpoint, identification of the gun that fired the bullet(s) requires that we compare the probabilities of observing matching striae under the competing hypotheses that the gun fired, or did not fire, the crime scene bullet. If indeed the uniqueness assumption is plausible, the latter probability approaches zero and the former approaches (but never reaches) one.

Current firearm examination practice relies mostly on visual assessment and comparison of striation. Indeed, the AFTE Theory of Identification (https://afte.org/about-us/what-is-afte/afte-theory-of-identification) explicitly requires that examiners evaluate the strength of similarity between two samples relative to other comparisons they may have carried out in the past. An attempt to quantify the degree of similarity consists in counting the number of consecutively matching striae (CMS) between two bullets, first proposed by \citet{biasotti:1959}. This approach has two drawbacks, however. First, determining matching striae is still a subjective activity. Second, as discussed by \citet{miller:1998}, the number of CMS may be high even if the bullets were not fired by the same gun.

Here, we focus on the question of defining a metric that can be used to objectively compare two bullets. We propose a framework which allows for the automatic analysis of the surface topologies of bullets, and the transcription of the individual characteristics into a set of features that quantify their similarities.
This allows for an objective and quantitative assessment of striae-based bullet matches.


We work with images from the James Hamby Consecutively Rifled Ruger Barrel Study~\citep{hamby:2009}. Ten consecutively rifled Ruger P-85 pistol barrels were obtained from the manufacturer and fired to produce 20 known test bullets and 15 unknown bullets for comparison.
3D topographical images of each bullet were obtained using a NanoFocus lens at 20x magnification and made publicly available on the NIST Ballistics Database Project\footnote{\url{http://www.nist.gov/forensics/ballisticsdb/hamby-consecutively-rifled-barrels.cfm}} in a format called x3p (XML 3-D Surface Profile). The x3p format conforms to the ISO5436-2 standard\footnote{\url{http://sourceforge.net/p/open-gps/mwiki/X3p/}} and is implemented to provide a simple and standard conforming way to exchange 2D and 3D profile data. It was adopted by the OpenFMC (Open Forensic Metrology Consortium\footnote{\url{http://www.openfmc.org/}}), a group of academic, industry, and government firearm forensics researchers whose aim is to establish best practices for researchers using metrology in forensic science. We have developed an open-source package for analyzing bullet land impressions written in R \citep{R}. This package is called bulletr \citep{bulletr} and enables direct reading and manipulation of x3p files. It also implements all of the methods we propose in this paper. A different package exists for reading x3p files called x3pr \citep{x3pr}, developed by Petraco (2014), but it is not designed to carry out calculations like the ones we propose after the x3p files have been read.

\begin{figure}[hbtp]
  \centering
\begin{subfigure}[t]{\textwidth}\centering
\caption{View of the data along the circumference of the bullet (circular segment of about 30 degrees).\label{fig:sidex3p}}{%
      \includegraphics[width=.65\textwidth]{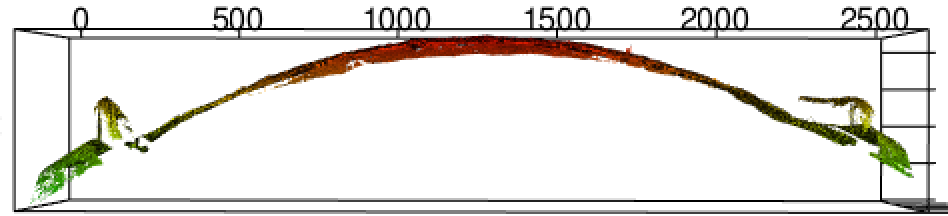}
      \hspace{1cm}
      \includegraphics[width=.25\textwidth]{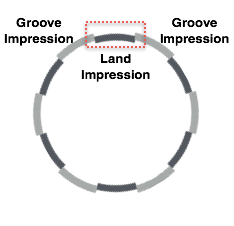}
    }
\end{subfigure}
\begin{subfigure}[t]{\textwidth}\centering
    \caption{Frontal view of a bullet land impression (lower end of the view is the bottom of the bullet). \label{fig:topx3p}}{%
    \includegraphics[width=.65\textwidth]{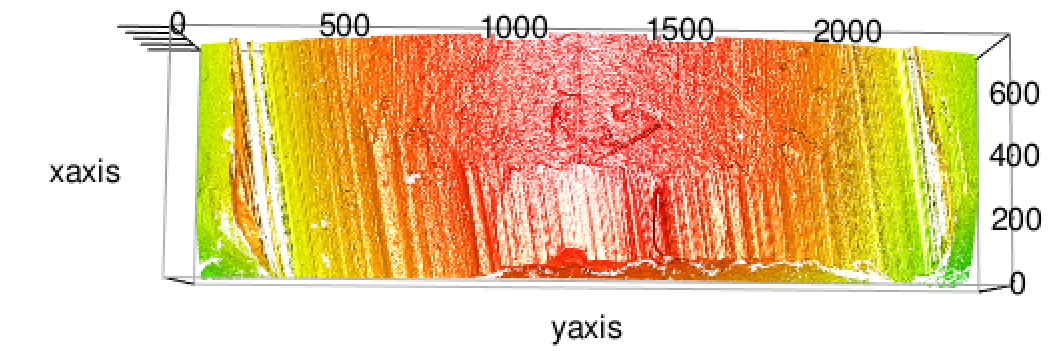} \hspace{1cm}
    \includegraphics[width=.15\textwidth]{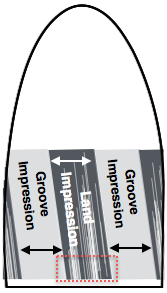}
    }
\end{subfigure}
\caption{Example of a groove-to-groove scan of a bullet land impression. The red-dotted rectangle on the right shows the location and orientation of the segment. }
\label{fig:allx3p}
\end{figure}

Each fired bullet is provided in the form of a set of six x3p files, where each file is a surface scan between adjacent groove impressions on the bullet, called a ``land" or land impression. The shoulders are the raised portion of the surface between the groove impression and the land impression. In the Hamby data, typical length (shoulder-to-shoulder) of a land impression is about 1998.28 micrometers or 2 millimeters. For notational simplicity, we refer to a particular land impression of a bullet as bullet X-Y, where X is the bullet identifier, and Y is the land number. An example of plotting one of these land impressions is given in Figures~\ref{fig:sidex3p} and~\ref{fig:topx3p}. These figures show side and top profiles of the land respectively. The tilt of the lines to the left in Figure~\ref{fig:topx3p} is not an artifact, but a direct and expected consequence of the spin induced by the rifling during the firing process. Depending on whether a barrel is rifled clockwise or counter-clockwise, the striations have a left or right tilt. The direction of the rifling is a class characteristic, i.e.\ a feature that pertains to a particular class of firearms, and is not unique at the individual barrel/bullet level.

The typical number and width of striation markings on bullets varies significantly depending on the gun barrel. For instance, a Smith and Wesson barrel with a land-width of 2.4 millimeters contained an average 60 striae, with an average width of about 0.08 millimeters~\citep{chu:2011}.

The purpose of our paper is to present an automatic matching routine that allows for a completely objective assessment of the strength of a match between two bullet land impressions. While we assess the performance of the algorithm in terms of a binary decision of match vs. non-match using a 50\% probability cut-off, our primary goal is to highlight the features that are statistically associated with matches and non-matches, and to provide a quantitative assessment of this association. In a real-world application of our algorithm, the raw scores would need further analysis and scrutiny, and it is likely that a 50\% cut-off would be an inappropriate choice on the basis of reasonable doubt.

Our algorithm is fully open source and available on GitHub \citep{bulletr}. This transparency allows for a greater understanding of the individual steps involved in the bullet matching process, and allows other forensic examiners, as well as outside observers, to examine the factors that discriminate between known bullet matches and non-matches. We have chosen to perform the matching on a land-to-land level, rather than bullet-to-bullet level. Although doing so introduces an implicit assumption of independence between land impressions, assuming independence only serves to make the task more challenging.

The remainder of this paper is structured as follows: We first briefly review some earlier work. We then discuss two methods for modeling the class structure of the bullet surfaces. Finally, we proceed to describing an automatic matching routine which we evaluate on the bullets made available through the Hamby study.

\section{Previous work}

There have been attempts to develop automatic or semi-automatic matching protocols, but most have focused on breech face and firing pin marks \citep[e.g.][]{riva:2014} or discuss a single attribute for comparison \citep[e.g.][]{vorburger:2011, chu:2011}. Still others refer to proprietary algorithms \citep{roberge:2006}. We briefly review some of this earlier work in what follows.

The original paper on the complete Hamby study already reports the successful use of several computer-assisted methods. However, aside from a zero false positive rate, false-negative error rates for bullets are not given nor are error rates for land-to-land matches mentioned.

\citet{lock2013significance} proposed an approach to quantify similarity of toolmarks. Their algorithm determines an optimal matching window between two toolmark signatures, and then performs a set of both coordinated and independent shifts. Given a match, the coordinated shifts would be expected to yield correlation values higher than those obtained from independent shifts. This is assessed using a Mann-Whitney U Statistic.

A procedure for bullet matching using the BulletTrax3D system is described in \citet{roberge:2006}. Their study used a different set of ten consecutively rifled barrels; matches are identified based on a bullet-to-bullet correlation score. The authors state that this process `could be automated', but no implementation of the algorithm is available.

Modern automated techniques using 3D images have also been proposed by \citet{riva:2014}. However, the authors focused on cartridge cases and not bullets. This might seem like a trivial distinction, but it has implications for the development of the algorithm. Their algorithm performs alignment of striae by rotation of the XY plane, which is not generalizable to bullets in which the XY plane is not flat.

Other work on 3D images has been described by \citet{petraco:2012}, who also focus on cartridge cases, as well as screwdriver striation patterns, and by others \citep[e.g.,][]{chu:2010, chu:2011, vorburger:2011}.

\section{Bullet signatures}

\begin{figure}[hbtp]
  \centering
\begin{knitrout}
\definecolor{shadecolor}{rgb}{0.969, 0.969, 0.969}\color{fgcolor}
\includegraphics[width=0.65\textwidth]{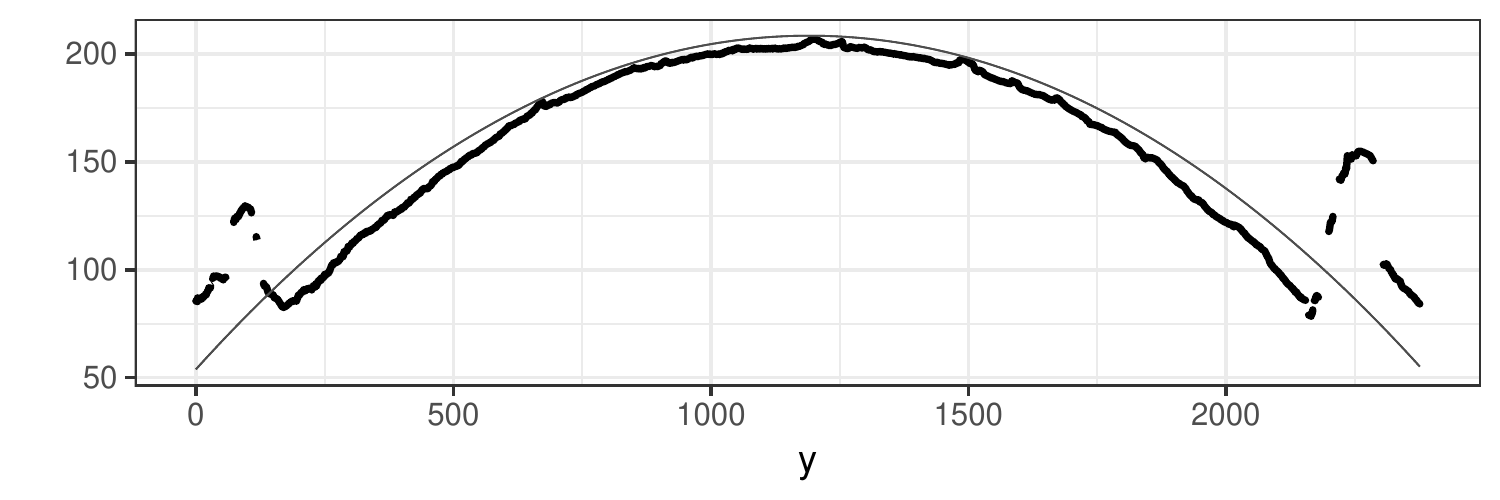}

\end{knitrout}
\caption{\label{fig:fixedX}Side profile of the surface measurements (in~$\mu m$) of a bullet land impression at a fixed height $x$. Note that the global features dominate any deviations, corresponding to the individual characteristics of striation marks.}
\end{figure}

To analyze the striation pattern, we extract a \emph{bullet profile} \citep{ma:2004} by taking a cross section of the surface measurements at a fixed height $x$ along the bullet land impression, as previously illustrated in Figure \ref{fig:allx3p}.
Figure \ref{fig:fixedX} shows a plot of the side profile of a bullet land impression. It can be seen that the global structure of the land dominates the appearance of the plot. The shoulders can be clearly identified on the left and right side, and the curvature of the surface is the most visible feature in the middle.

The smooth curve on the plot represents a segment of a perfect circle with the same radius as the bullet. While the circle is an obvious first choice for fitting the structure, it does not completely capture the bullet surface after it was fired. A discussion of a circular fit and the remaining residual structure can be found in Supplement Section~\ref{supp:cylindrical}.

Instead of a circular fit, we use multiple loess fits to model the overall structure and extract the bullet markings.

\subsection{Identifying shoulder locations}
\begin{figure}[hbtp]
  \centering
\begin{subfigure}[t]{\textwidth}\centering
\caption{\label{fig:loess_step1}Step 1 of identifying shoulder locations: For a fixed height ($x = 243.75\mu m$)  surface measurements for bullet~1-5 are plotted across the range of $y$.}
\begin{knitrout}
\definecolor{shadecolor}{rgb}{0.969, 0.969, 0.969}\color{fgcolor}
\includegraphics[width=.5\textwidth]{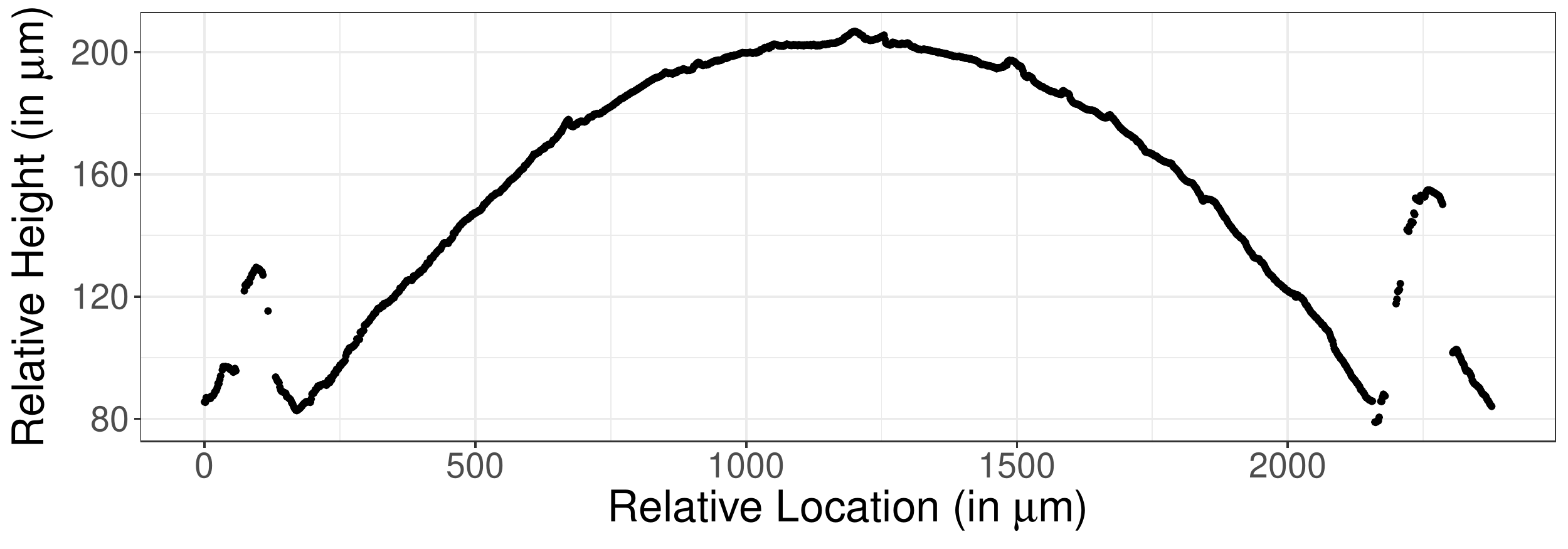}

\end{knitrout}
\end{subfigure}
\begin{subfigure}[t]{\textwidth}\centering
\caption{\label{fig:loess_step3}Step 2 of identifying shoulder locations: The  surface measurements are smoothed twice with a smoothing factor of $s = 35$. The orange rectangle shows an example of the smoothing window. Valleys and peaks are detected, if they are not within the same window.}
\begin{knitrout}
\definecolor{shadecolor}{rgb}{0.969, 0.969, 0.969}\color{fgcolor}
\includegraphics[width=.5\textwidth]{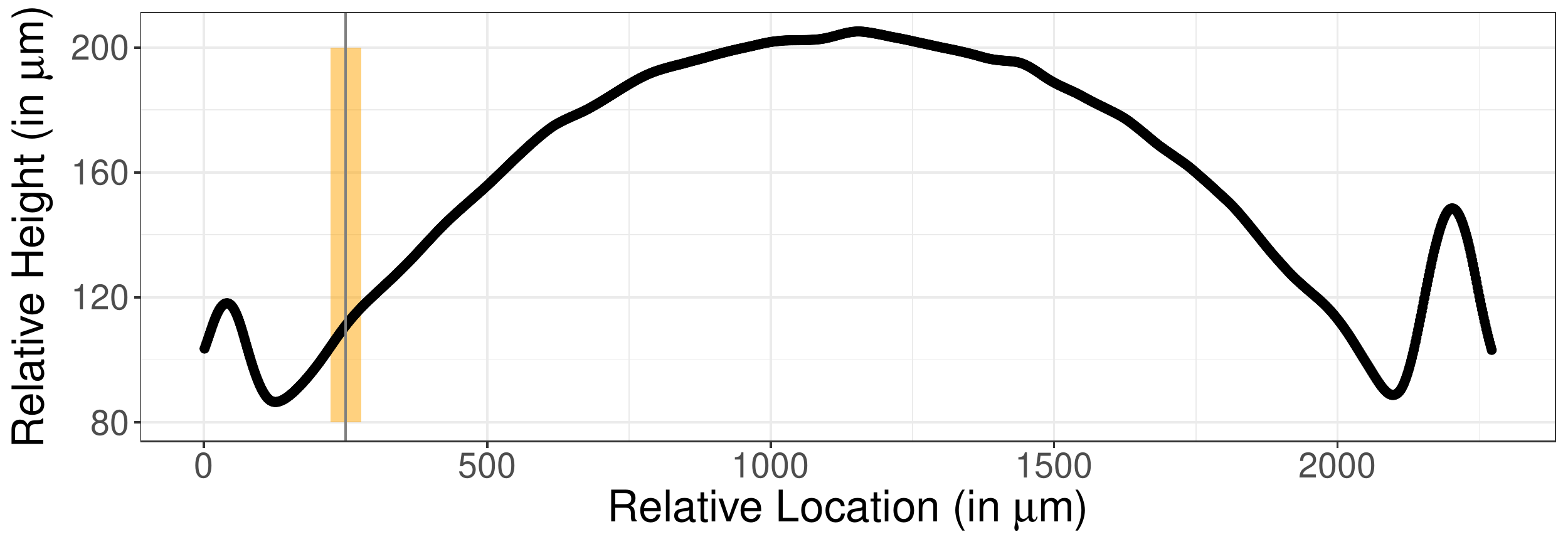}

\end{knitrout}
\end{subfigure}
\begin{subfigure}[t]{\textwidth}\centering
\caption{\label{fig:loess_step47}Steps 3 -- 6 of identifying shoulder locations: After smoothing the surface measurements extrema on the left and right are detected (marked by vertical lines, red indicating peaks and blue indicating valleys). Values outside the blue boundaries are removed (shown in grey)}
\begin{knitrout}
\definecolor{shadecolor}{rgb}{0.969, 0.969, 0.969}\color{fgcolor}
\includegraphics[width=.5\textwidth]{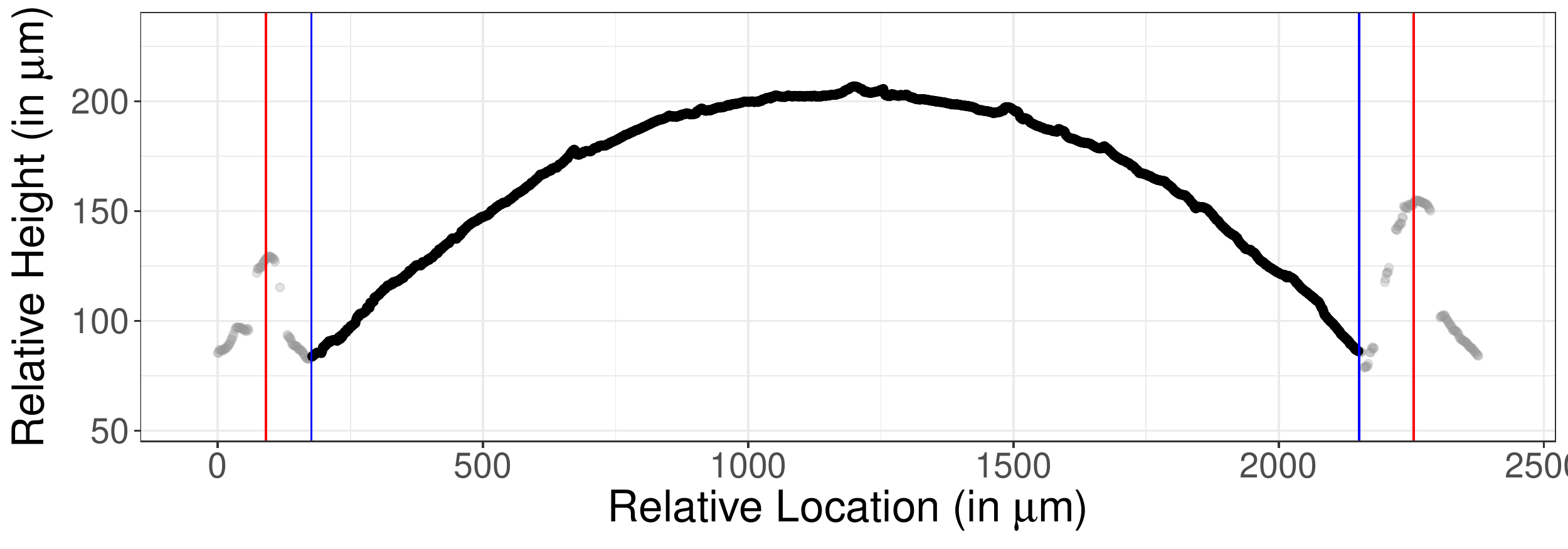}

\end{knitrout}
\end{subfigure}
\caption{Overview of all six steps of the smoothing algorithm to identify and remove shoulders and groove impressions from the bullet images.}
\end{figure}

We first identify the location of the left and right shoulders in the image. The groove impressions are assumed to contain no information relevant for determining matches, and the shoulders identify the location at which the land impression begins. The shoulders also dominate the structure, and therefore need to be removed.
Fortunately, the location and appearance of the shoulders in the surface profiles is quite consistent.
Surface measurements reach local maxima around the peak of the shoulder at either end of the range of $y$, and we can then follow the descent of the surface measurements inwards to the valley of the shoulder.
The location of the valleys mark the points at which we trim the image. The procedure can be described as follows:

\begin{enumerate}
    \item At a fixed height $x$ extract a bullet's profile (Figure~\ref{fig:loess_step1}, with $x = 243.75\mu m$).
    \item For each $y$ value, smooth out any deviations occurring near the minima by twice applying a rolling average with a pre-set \emph{smoothing factor} $s$. (Figure~\ref{fig:loess_step3}, smoothing factor $s = 35$(data points) corresponding to 55$\mu m$).
    \item Determine the location of the peak of the left shoulder by finding the first doubly-smoothed value $y_i$ that is the maximum within its smoothing window (e.g.\ such that $y_i > y_{i - 1}$ and $y_i > y_{i + 1}$, where $i$ is between 1  and $\lfloor s/2 \rfloor$). We call the location of this peak $p_{\ell}$ (see Figure~\ref{fig:loess_step47}).
    \item Similarly, determine the location of the valley of the left shoulder by finding the first double-smoothed $y_j$ that is the minimum within its smoothing window. Call the location of this valley $v_{\ell}$.
    \item Reverse the order of the $y$ values and repeat the previous two steps to find the peak and valley of the right shoulder, $(p_{r}, v_{r})$.
    \item Trim the surface measurements to values within the two shoulder valleys (i.e.\ remove all records with $y_i < v_{\ell}$ and $y_i > v_{r}$) (see Figure~\ref{fig:loess_step47}).
\end{enumerate}

The smoothing factor $s$ introduced in the algorithm represents the window size to use for a rolling average. Higher values of $s$ therefore lead to more smoothing. Empirically, a value of $s = 35$ for the smoothing factor seems to work well (the smoothing factor is further discussed in Section~\ref{sec:smoothing}). It is important to note that the smoothing pass is done twice. That is, the smoothed data are once again smoothed by computing a new rolling average with the same smoothing factor. This bears some similarities to the ideas of John Tukey in his book Exploratory Data Analysis, where he describes a smoothing process called ``twicing" in which a second pass is made on the residuals computed from the first pass and then added back to the result~\citep{tukey:1977}. This has the effect of introducing a bit more variance back into the smoothed data.
We instead performed a second smoothing pass on the smoothed data, which has the effect of weighting observations near the center of the window the highest, with the weights linearly dropping off as we reach either end of the smoothing window.

\subsection{Removing curvature}
Next, we fit a loess regression to the data. Loess regression \citep{cleveland:1979} is based on the assumption that the relationship between two random variables $X$ and $Y$ can be described in the form of a smooth, continuous function $f$ with $y_i = f(x_i) + \varepsilon_i$ for all values $i = 1, ..., n$. The function $f$ is approximated via locally weighted polynomial regressions. Parameters of the estimation are $\alpha$, the proportion of all points included in the fit (here, $\alpha = 0.75$), the weighting function and the degree of the polynomial (here, we fit a quadratic regression).

The main idea of locally weighted regression is to use a weighting routine that emphasizes the effect of points close to the fitting location and de-emphasizes the effect of points as they are further away. The weighting function used here is the tricubic function $w(d) = \left(1 - d^3\right)^3$, for $d \in [0,1]$ and $w(d) = 0$ otherwise. Here, $d$ is defined as the distance between $x_i$ and the location of the fit $x_o$, divided by the overall range of the included data, so as to map the results to a $[0,1]$ range as the definition requires.

\begin{figure}[hbtp]
  \centering
\begin{subfigure}[b]{.49\textwidth}\centering
\caption{\label{fig:loess_fit} Loess fit for bullet~1-5.}
\begin{knitrout}
\definecolor{shadecolor}{rgb}{0.969, 0.969, 0.969}\color{fgcolor}
\includegraphics[width=\textwidth]{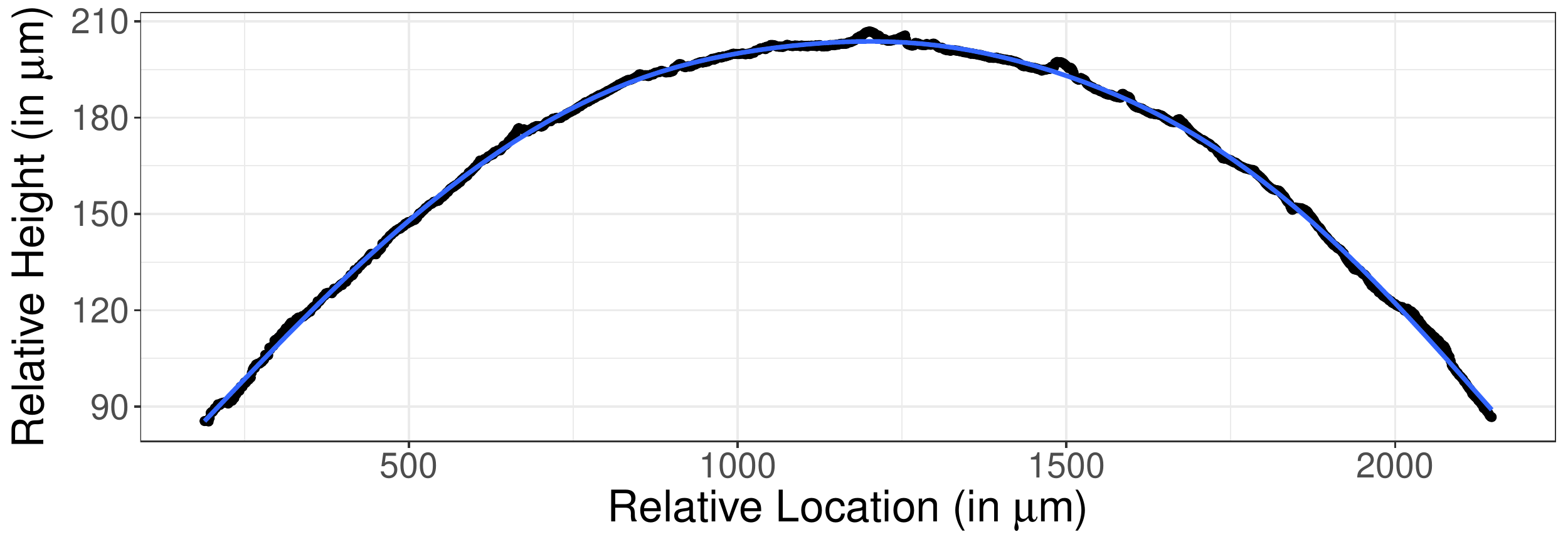}

\end{knitrout}
\end{subfigure}
\begin{subfigure}[b]{.49\textwidth}\centering
\caption{\label{fig:loess_resid} Residuals of loess fit for bullet~1-5.
}
\begin{knitrout}
\definecolor{shadecolor}{rgb}{0.969, 0.969, 0.969}\color{fgcolor}
\includegraphics[width=\textwidth]{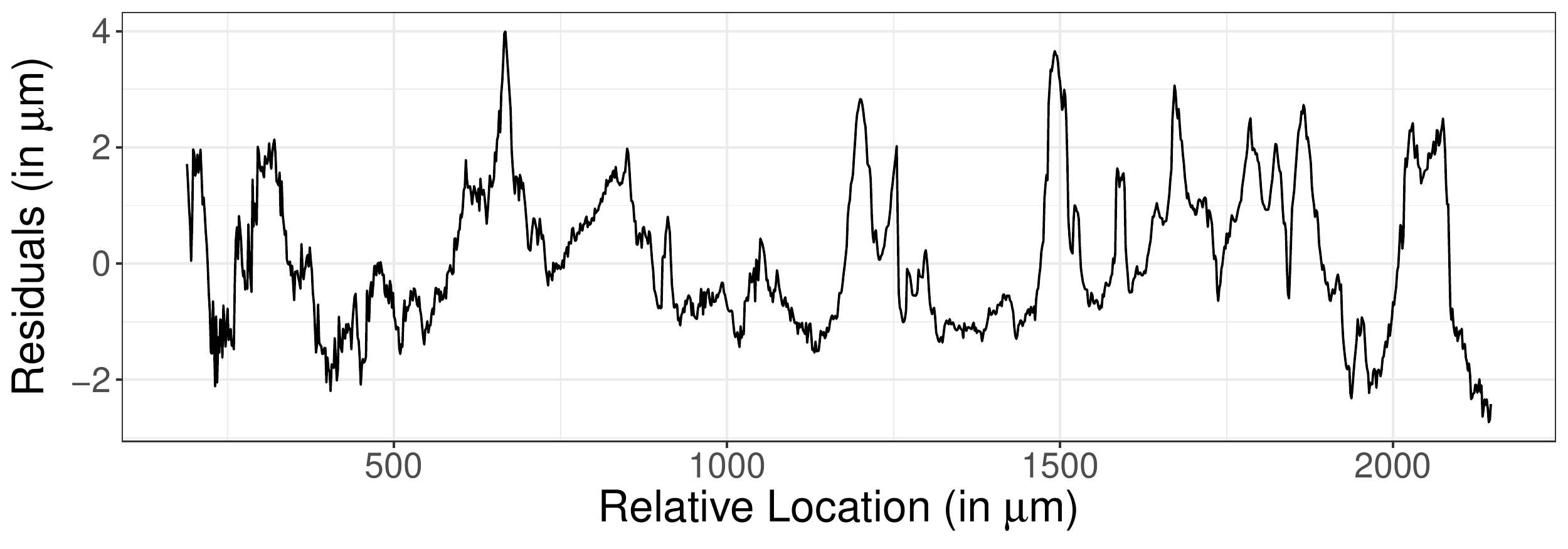}

\end{knitrout}
\end{subfigure}
\caption{Fit and residuals of a loess fit to bullet~1-5 (Barrel~1). The residuals define the {\it signature} of bullet~1-5. 
}
\end{figure}

Figure~\ref{fig:loess_fit} shows the loess fit, in blue, overlaid on the processed image of bullet~1-5. The fit seems to do a reasonable job of capturing the structure of the image. 
Figure~\ref{fig:loess_resid} shows the residuals from this fit. These residuals are called the \emph{signature} of bullet~1-5.

\section{Automatic matching}
\begin{figure}[hbtp]
\centering
\includegraphics[width=.65\columnwidth]{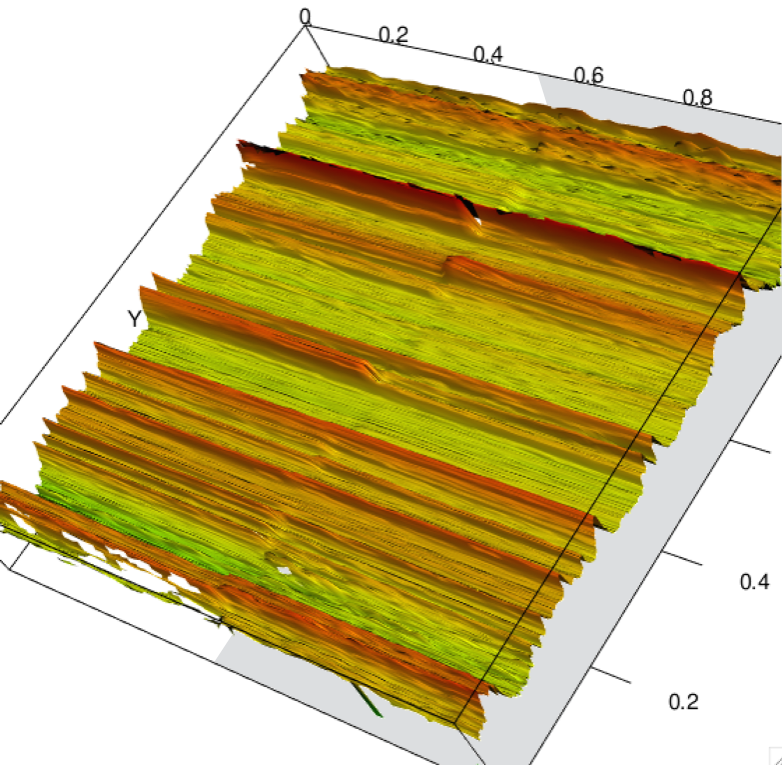}
\caption{\label{fig:manualmatch-rgl}3D view of the manually adjusted side-by-side comparison of bullet~1-5 and bullet 2-1 after removing the curvature. Bullet 2-1 is shaded light grey in the background.}
\end{figure}

Applying the loess fit to a range of different signatures (see Figure~\ref{fig:manualmatch-rgl} for signatures extracted at heights between 50$\mu m$ and 150~$\mu m$) shows the 3D striation marks from two bullets. Signatures of bullet~1 are shown on the left (all extracted from heights below 100$\mu m$) and signatures of bullet 2 are shown on the right (extracted at heights above 100$\mu m$). Signatures are manually aligned, resulting in many of the striation marks to continuously pass from one side to the other. Visually, this allows for an easy assessment of these two bullet land impressions as a match. However, this match relies on visual inspection and is therefore subjective.  The goal of this section is to eliminate the need for a visual inspection during the matching process and replace it by an automatic algorithm. This also allows for a quantification of the strength of the match.

In this section, we describe the algorithm for matching signatures first, and the impact of parameter choices in the subsections thereafter.

\subsection{Algorithm}\label{sec:algorithm}

\begin{figure}[hbtp]
\centering
    \begin{subfigure}[t]{\textwidth}\centering
\caption{Loess smooth of signatures  at a height of $x = 100\mu m$ (span is 0.03).\label{fig:smooth}}{%
\begin{knitrout}
\definecolor{shadecolor}{rgb}{0.969, 0.969, 0.969}\color{fgcolor}
\includegraphics[width=.65\textwidth]{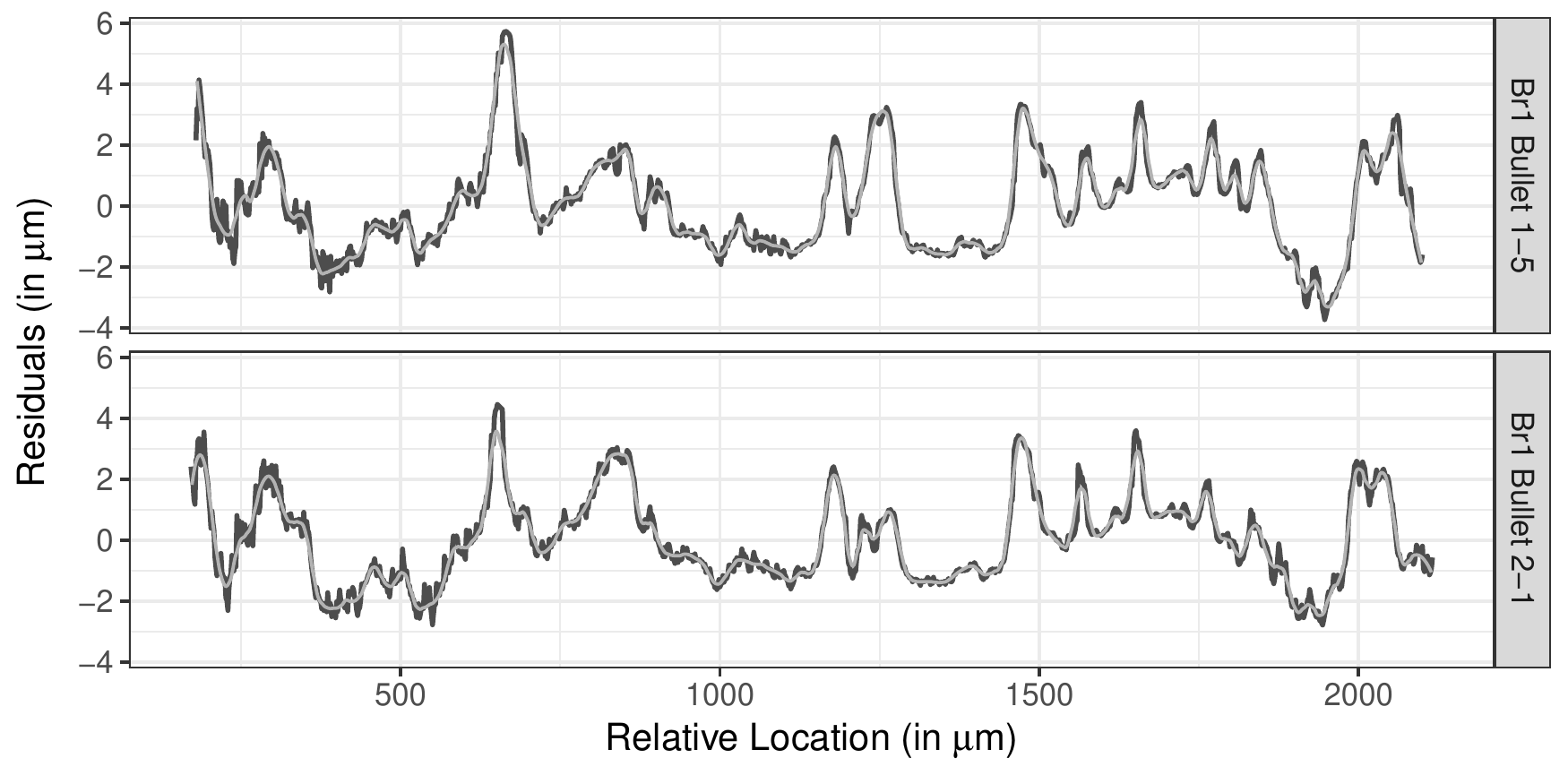}

\end{knitrout}
    }
\end{subfigure}
\begin{subfigure}[t]{\textwidth}\centering
\caption{Using a rolling median peaks and valleys are identified for each signature. Peaks and valleys on the signature correspond to striation marks on the bullet's surface. \label{fig:smoothcutb}}{%
\begin{knitrout}
\definecolor{shadecolor}{rgb}{0.969, 0.969, 0.969}\color{fgcolor}
\includegraphics[width=.65\textwidth]{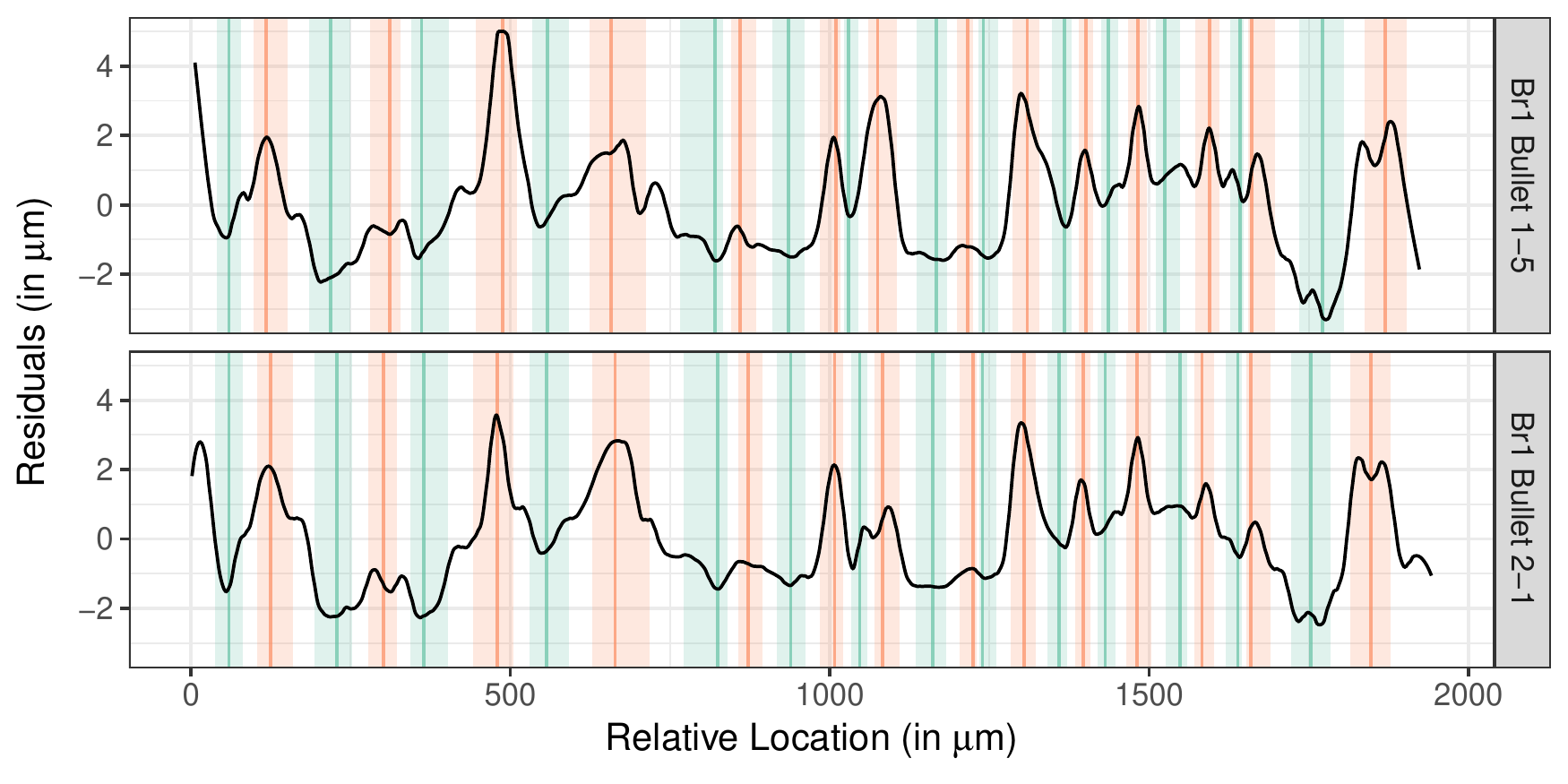}

\end{knitrout}
}
\end{subfigure}
\begin{subfigure}[t]{\textwidth}\centering
    \caption{Rectangles in the back identify a striation mark on one of the bullets.  Matching striation marks are indicated by color filled rectangles and marked by an `o'. Mismatches are filled in grey and  marked by an `x'.   \label{fig:smoothcutd}}{%
\begin{knitrout}
\definecolor{shadecolor}{rgb}{0.969, 0.969, 0.969}\color{fgcolor}
\includegraphics[width=.65\textwidth]{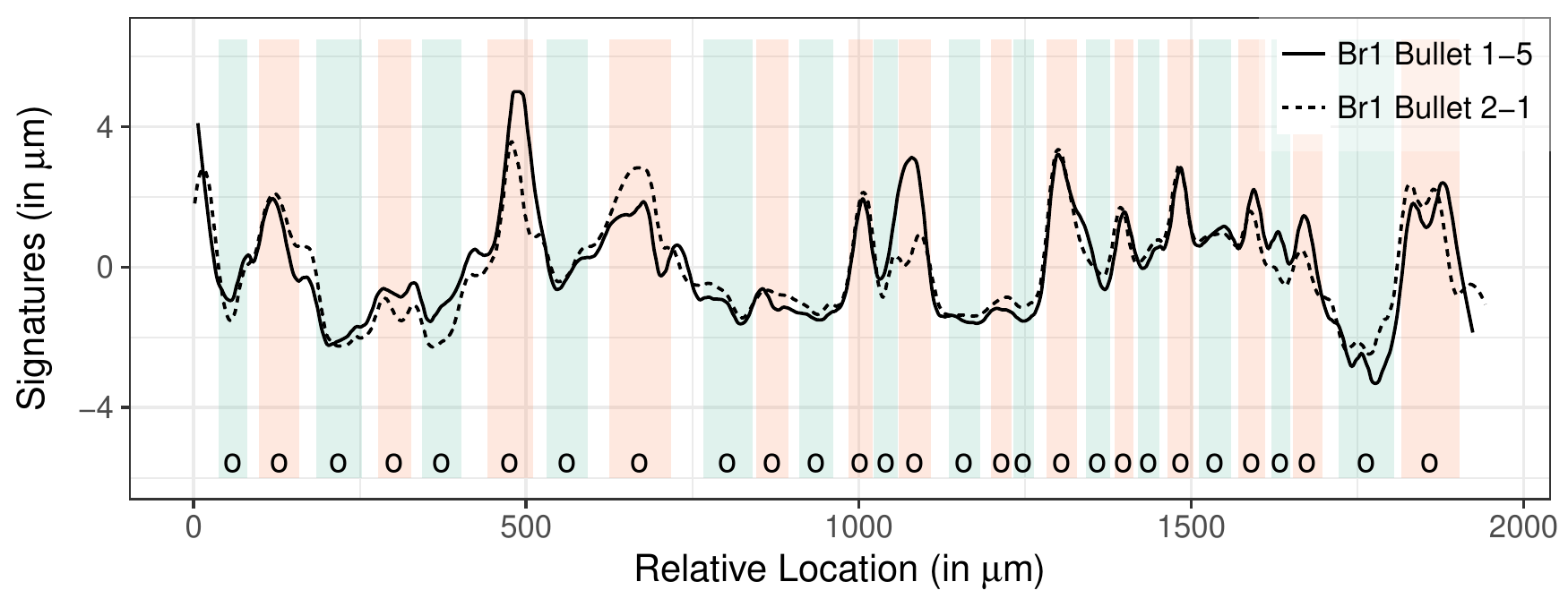}

\end{knitrout}
    }
\end{subfigure}
\caption{\label{fig:match}Matching striation marks: smooth (a), identify peaks and valley (b), and match peaks and valleys between signatures (c).}
\end{figure}
Figure~\ref{fig:match} gives an overview of the automated matching routine:
We first identify a stable region for each bullet land and extract the signature at the lowest height in this region, because typically, individual characteristics are best expressed at the lower end of the bullet (see Supplement Section~\ref{supp:bulletbottom} for a more detailed discussion).

All of the other steps are done on pairs of bullet land impressions:
\begin{enumerate}
\item {\bf Smooth the two signatures} using a loess with a very small span (see Figure~\ref{fig:smooth}).
\item Use cross-correlation to {\bf find the best alignment} of the two signatures: shift one of the signatures by the lag indicated by the cross-correlation function (see Figure~\ref{fig:ccf} for the cross-correlation function and Figure~\ref{fig:crosscutX} for the resulting shift).
\item Using a rolling average, {\bf identify peaks and valleys} for each of the signatures by identifying points at which the derivative of the signature is equal to zero. We then define an interval around the location of the extrema on each side as one third of the distance to the location of the next extrema (see Figure~\ref{fig:smoothcutb}). Peaks and valleys constitute the \emph{striation marks} on the bullet.
\item {\bf Match striations across signatures:} based on the intervals around the extrema as defined above, we identify common intervals as the areas in which two or more of the individual intervals overlap: a joint interval is defined as the smallest interval that encompasses all of the overlapping intervals. A joint interval is then called a match(ing stria) between the signatures, if all of the intervals are of the same type of extrema, i.e.\ they are either all peaks or all valleys. In Figure~\ref{fig:match} all matches are shown as color-filled rectangles corresponding to their type of extrema (peaks are shown in orange, and valleys in green). Non-matching intervals are left grey.
\item {\bf Extract features from the aligned signatures and the matches between them:} many different features can be extracted from the aligned signatures. Here, we describe a few of the ones that can be found in the literature and some that we found to be of practical relevance:
\begin{enumerate}[label=(\roman*)]
\item Maximal number of CMS (consecutive matching striae), and, similarly, the number of consecutively non-matching striae (CNMS),
\item Number of matches and non-matches,
\item The value of the cross-correlation function (ccf) between the aligned signatures \citep{vorburger:2011},
\item Average difference $D$ between signatures, defined as the Euclidean vertical distance between surface measurements of aligned signatures. Let $f(t)$ and $g(t)$ be smoothed, aligned signatures:
\[
D^2 = \frac{1}{\text{\#}t}\sum_t \left[f(t) - g(t)\right]^2,
\]
\item The sum $S$ of average absolute heights of matched extrema: for each of the two matched stria, compute the average of the absolute heights of the peaks or valleys. $S$ is then defined as the sum of all these averages.
\end{enumerate}
\end{enumerate}
The difference $D$ between signatures is here defined as the Euclidean distance (in $\mu m$). In the paper by \citet{ma:2004}, distance is defined as a measure relative to the first signature, which serves as a comparison reference and is therefore a unitless quantity.

Counting the maximal number of CMS is part of the current practice to identify bullet matches~\citep{nichols:1997, nichols:2003, nichols:2003b}.
In the example of Figure~\ref{fig:match}, the number of consecutive matching striations (CMS) is fifteen, a high number suggestive of a match between the land impressions. Note that the definition of CMS we use does not match the one given in \citet{thompson:2013}. There, CMS is defined only in terms of matching peaks without regarding valleys. Additionally, peaks in  \citet{thompson:2013} are  used only if they can be identified and matched `within a tolerable range' between land impressions. The definition given here is computationally less complex, but should yield highly correlated values, because of the requirement to only consider signatures from a stable region in the land (see Section~\ref{sec:heights} for further details on stability of regions). In the Hamby study, the definition of CMS by \citet{thompson:2013} leads to approximately half of the values of CMS defined in this paper (with a correlation coefficient between the values of the two definitions of about 0.92).
For lead bullets, such as used in the Hamby study, \citet{biasotti:1959} considered four or more consecutive peaks (corresponding to eight or more consecutive lines in our definition) to be sufficient evidence of a match.  

Determining a threshold such that CMS values above the threshold indicate a match with high reliability is beyond the scope of this work, even though it is critically important in practice. We provide some ideas in the next section, but first we assess the robustness of the matching algorithm to different choices of the parameter values.

\subsection{Horizontal alignment}
\begin{figure}[hbtp]
\centering
\begin{subfigure}[b]{.49\textwidth}\centering
\caption{Raw bullet land impression signatures.\label{fig:crosscut}}{%
\begin{knitrout}
\definecolor{shadecolor}{rgb}{0.969, 0.969, 0.969}\color{fgcolor}
\includegraphics[width=\textwidth]{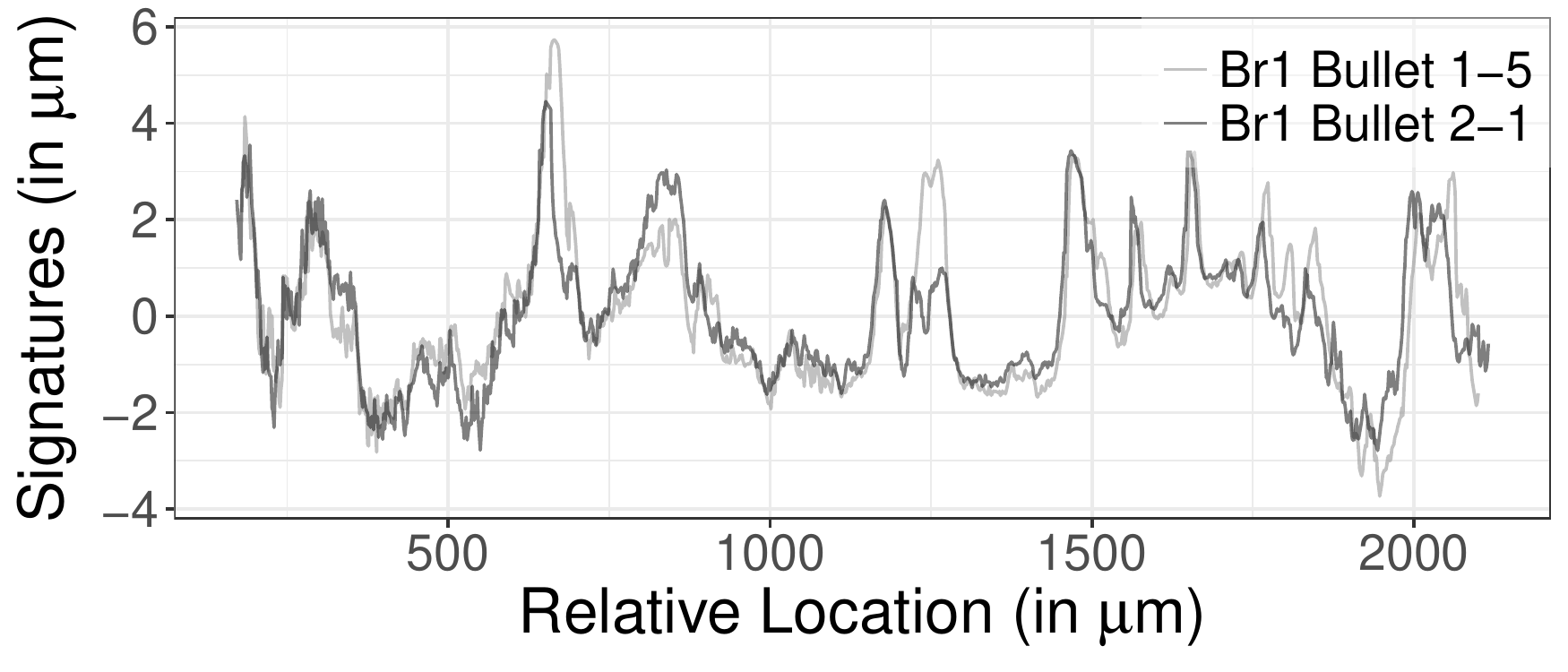}

\end{knitrout}
    }
\end{subfigure}
\begin{subfigure}[b]{.49\textwidth}\centering
    \caption{Aligned signatures.\label{fig:crosscutX}}{%
\begin{knitrout}
\definecolor{shadecolor}{rgb}{0.969, 0.969, 0.969}\color{fgcolor}
\includegraphics[width=\textwidth]{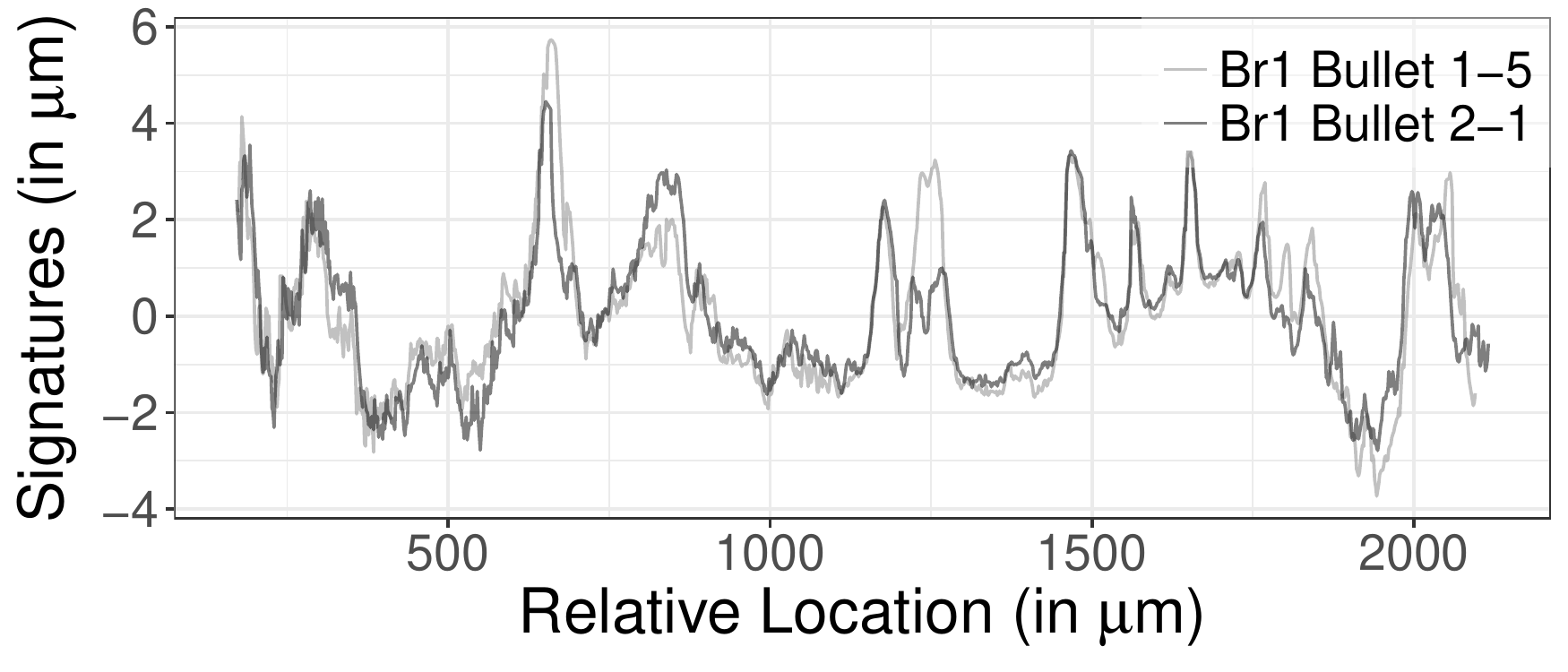}

\end{knitrout}
    }
\end{subfigure}
\caption{\label{fig:cross100}Signatures of bullets 1-5 and 2-1 taken at  heights of $x = 100\mu m$. A horizontal shift of the values of bullet~1-5 to the right shows the similarity of the striation marks.}
\end{figure}

\begin{figure}[hbtp]
  \centering
\begin{knitrout}
\definecolor{shadecolor}{rgb}{0.969, 0.969, 0.969}\color{fgcolor}
\includegraphics[width=.65\textwidth]{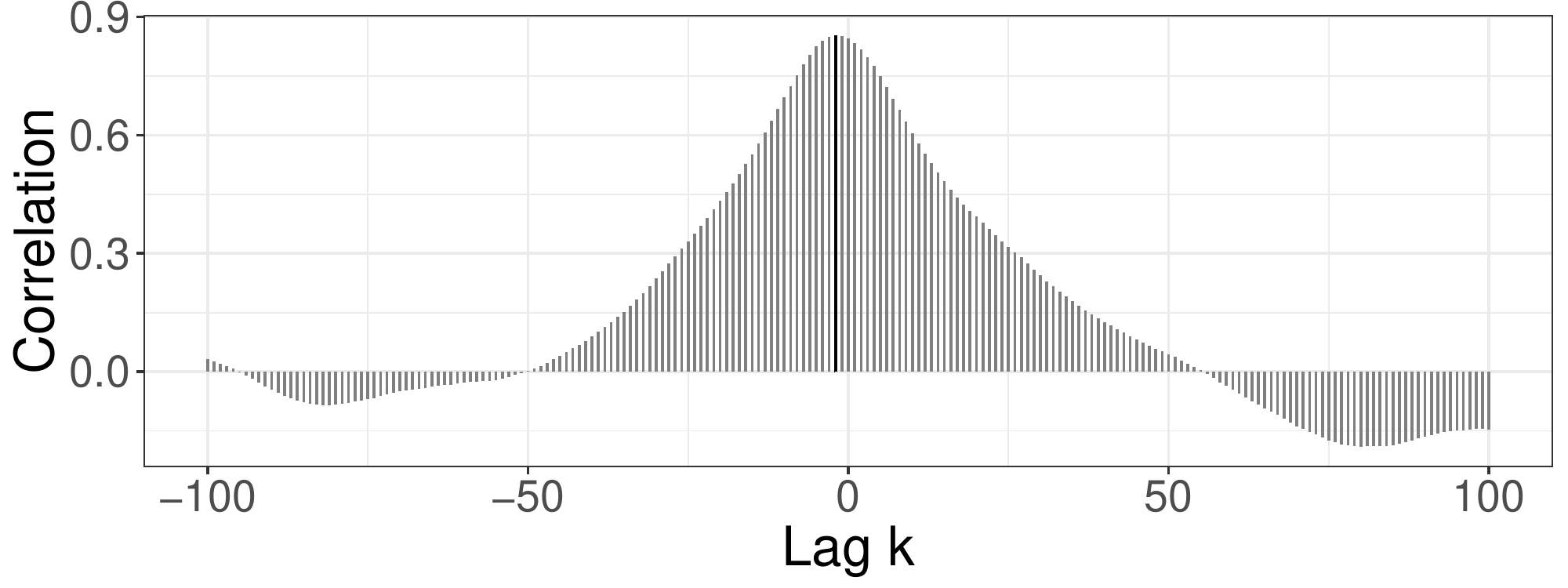}

\end{knitrout}
\caption{\label{fig:ccf}Cross-correlation function between the two signatures shown in Figure~\ref{fig:crosscut} at lags between -100 and 100. The correlation is maximized at a lag of -2, indicating the largest amount of agreement between the signatures. Figure~\ref{fig:crosscutX} shows the slight change resulting from the lag-shifted signatures.}
\end{figure}

Signatures of each of the two land impressions, 1-5 and 2-1,  in Figure~\ref{fig:manualmatch-rgl} are shown in Figure~\ref{fig:cross100} extracted at a height of $x = 100\mu m$. Striation marks show up in these representations as peaks and valleys.  The individual characteristics are prominent and, again, suggest a match between the land impressions. A horizontal shift of one of the signatures (result shown in Fig~\ref{fig:crosscutX}) emphasizes the strong similarities between signatures.
For this alignment we use the cross-correlation function to find a maximal amount of agreement between the signatures \citep{bachrach:2002, chu:2010, vorburger:2011, thompson:2013}.
This horizontal shift is based on the cross-correlation between the two signatures: let $f(t)$ and $g(t)$ define the signature values  at $t$, where $t$ are locations between 0~$\mu m$ and about 2500~$\mu m$, 1.5625~$\mu m$ apart.
The cross-correlation between $f$ and $g$ at lag $k$ is then defined as
\[
(f * g) (k) = \sum_t f(t+k) g(t),
\]
with suitably defined limits for the summation. 

\subsection{Impact of bullet height}\label{sec:heights}
\begin{figure}[hbtp]
  \centering
\begin{knitrout}
\definecolor{shadecolor}{rgb}{0.969, 0.969, 0.969}\color{fgcolor}
\includegraphics[width=\textwidth]{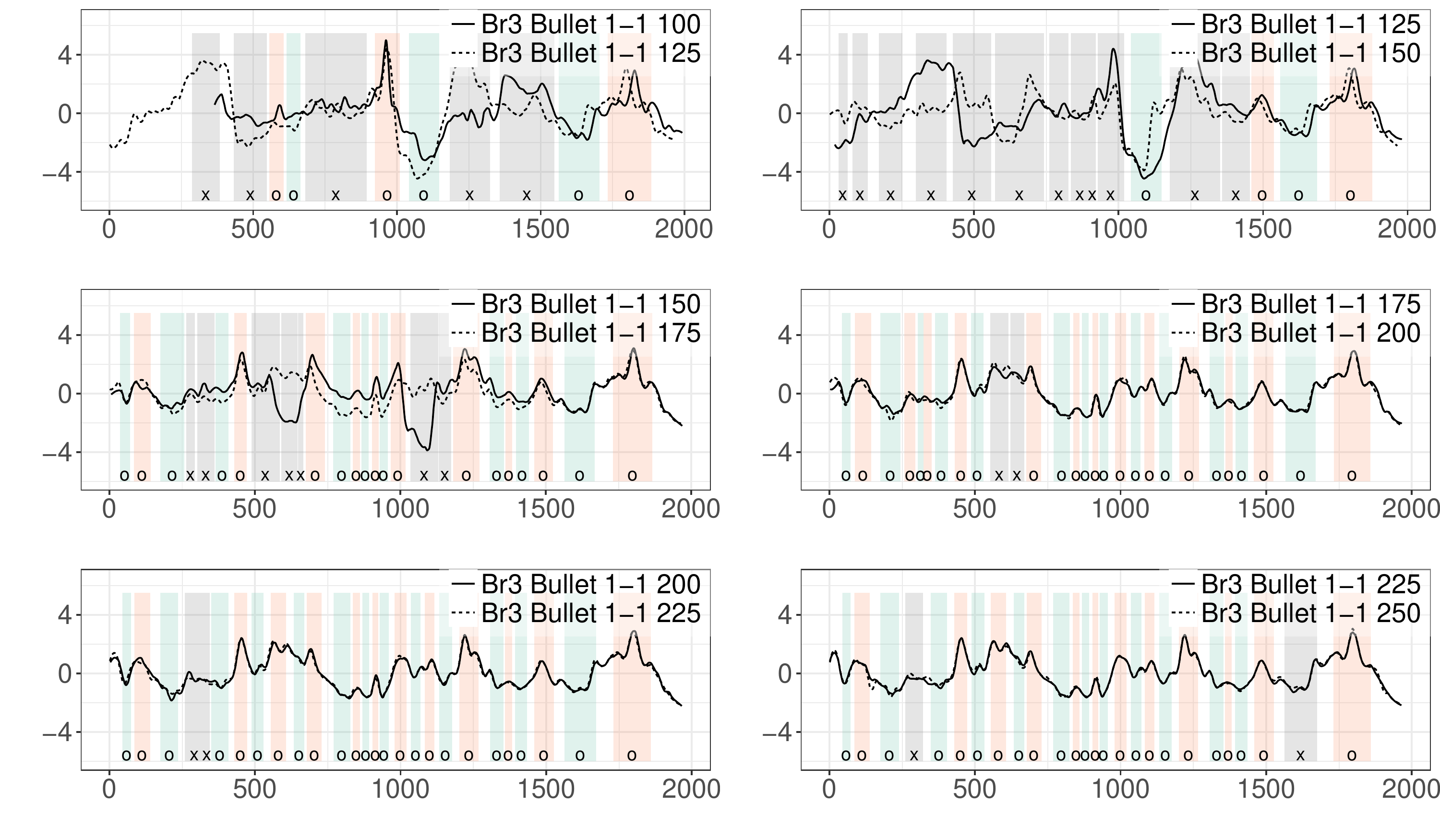}

\end{knitrout}
\caption{\label{fig:crosscuts2}Signatures  for barrel 3, bullet~1-1 extracted from varying heights. Initially, the match between signatures taken at heights 25$\mu m$ apart is affected strongly by some break off at the bottom of the bullet. At a  level of $175\mu m$ the bullet's signature stabilizes. For this land impression, matches should not be attempted at lower heights. }
\end{figure}

The height at which signatures are extracted for a comparison between bullet land impressions matters -- signatures taken from heights that are further apart, show more pronounced differences between the signatures.
This poses both a caveat to matching attempts as well as an opportunity for quality control: we have to be aware of the height that was used in a matching. Visually, matches degrade if the signatures upon which the match is based are from heights further than 200$\mu m$ apart  (see Supplement Section~\ref{supp:ccf} for more discussion). However, we can extract signatures from multiple heights of the same bullet land impression for an initial assessment of its quality. By comparing signatures from heights that are not too far apart --  25~$\mu m$ to 50~$\mu m$ -- we get an indication whether the signatures come from a rapidly changing section of the surface, indicative of a break-off or some other damage, or from a stable section, where we have a reasonable expectation of finding matches to other signatures. In the approach here, we keep increasing the height $x$ at which the signature is taken until we find a section with a stable pattern. This process is shown in Figure~\ref{fig:crosscuts2} at the example of bullet~1-1 from barrel 3, where `stability' is  defined as two aligned signatures from heights chosen 25$\mu m$ apart having a cross-correlation of at least 0.95.

\subsection{Varying smoothing factor}\label{sec:smoothing}
As mentioned earlier, the algorithm for detecting peaks and valleys depends on the selection of a smoothing window, called the smoothing factor or span. A smoothing factor of $k$ means that the  $k$ closest observations to $x_o$ are considered for a fit for $x_o$. Because surface measurements are recorded at equal sampling intervals (here, of 1.5625$\mu m$), we decided to only consider odd smoothing factors $2k + 1$, which means that the  $k$ observations to the left and right of $x_o$ are considered for a local fit of $x_o$. For detecting and removing the shoulders prior to fitting a loess regression we selected a smoothing factor of 35, while for detecting the peaks/valleys of the loess residuals a smoothing factor of 25 seems more appropriate. 

\begin{figure}[hbtp]
  \centering
\begin{knitrout}
\definecolor{shadecolor}{rgb}{0.969, 0.969, 0.969}\color{fgcolor}
\includegraphics[width=.75\textwidth]{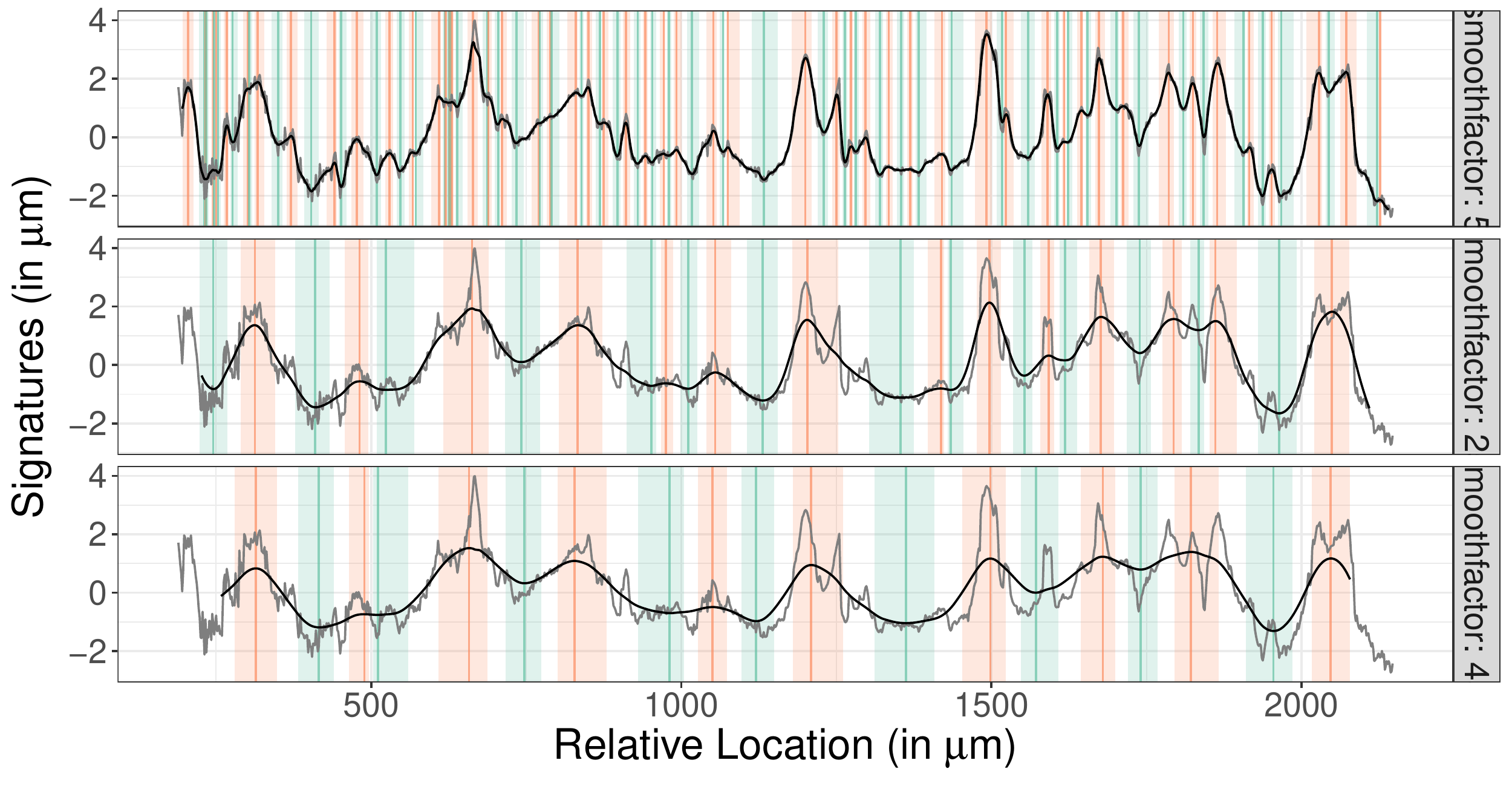}

\end{knitrout}
\caption{\label{fig:varysmooth} Peak/valley detection at smoothing factors of 5, 25, and 45, respectively. Note that a smoothing factor of 5 yields enough noise that many very minimal overlapping peaks and valleys are detected, while a smoothing factor of 45 might over-smooth and cause the peaks/valleys to either end disappear or shift  horizontally from their original position in the signature.}
\end{figure}

Figure~\ref{fig:varysmooth} displays the  peaks and valleys detected in the same signature at smoothing factors of 5, 25, and 45, respectively. The dark line corresponds to the smoothed values, while the grey line in the back shows the raw signature. The choice of smoothing factor is a classical decision of a bias/variance trade-off. It is immediately clear that a small smoothing factor like 5 is a poor choice. It results in a significant amount of noise in the data such that even just a point or two can skew the rolling average enough for a peak or valley to be detected. Given that striation widths are typically much larger, we are in effect muddying the waters by performing such minimal smoothing. Another consideration is that the smoothing should not fall below the  resolution of the equipment at which the surface measurements are taken -- so as to not introduce artifacts in the analysis.

A larger smoothing factor on the other hand (like 45), seems to be a more plausible option. Most of the peaks/valleys present which are detected by a smoothing factor of 25 are also detected at 45. However, some notable issues arise. Notice that the valley on the right hand side of the image is smoothed out, and thus not detected. On the left hand side, a double peak is detected - that might be a questionable decision - but there are several peaks in the middle, that are smoothed out, for example the peak at around $y = 750$. That is, in many cases, large windows are smoothing out some of the structure that we wish to see. Furthermore, it can be seen that the peaks/valleys are often shifted relative to their position in the original loess residuals, or in the smoothed data with smaller smoothing factors.

\section{Evaluation}
\begin{figure}
  \centering
\begin{subfigure}[t]{.49\textwidth}\centering
\caption{Barrel 6 Bullet 2-1}
\includegraphics[width=\textwidth]{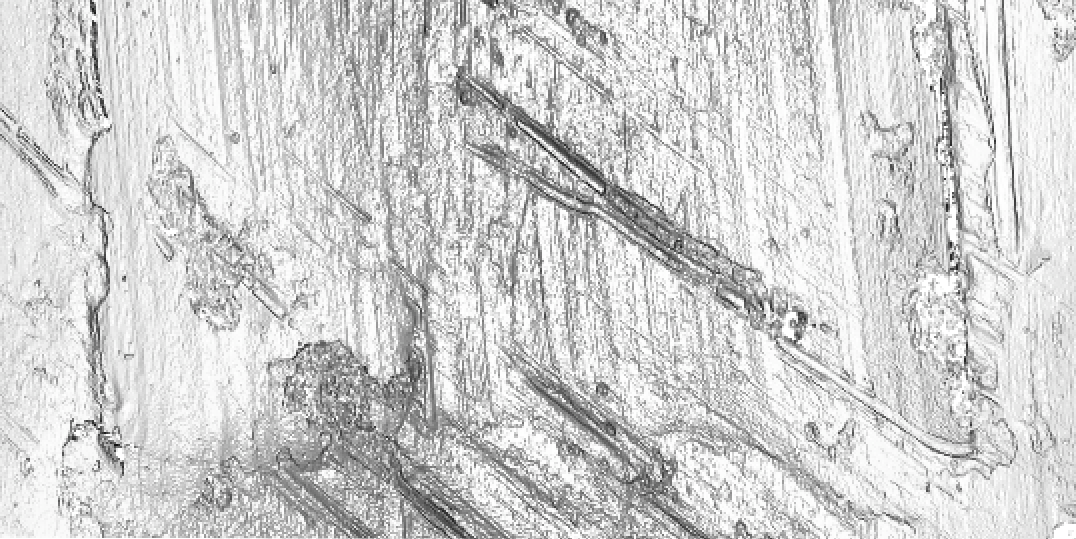}
\end{subfigure}
\begin{subfigure}[t]{.49\textwidth}\centering
\caption{Barrel 9 Bullet 2-4}
\includegraphics[width=\textwidth]{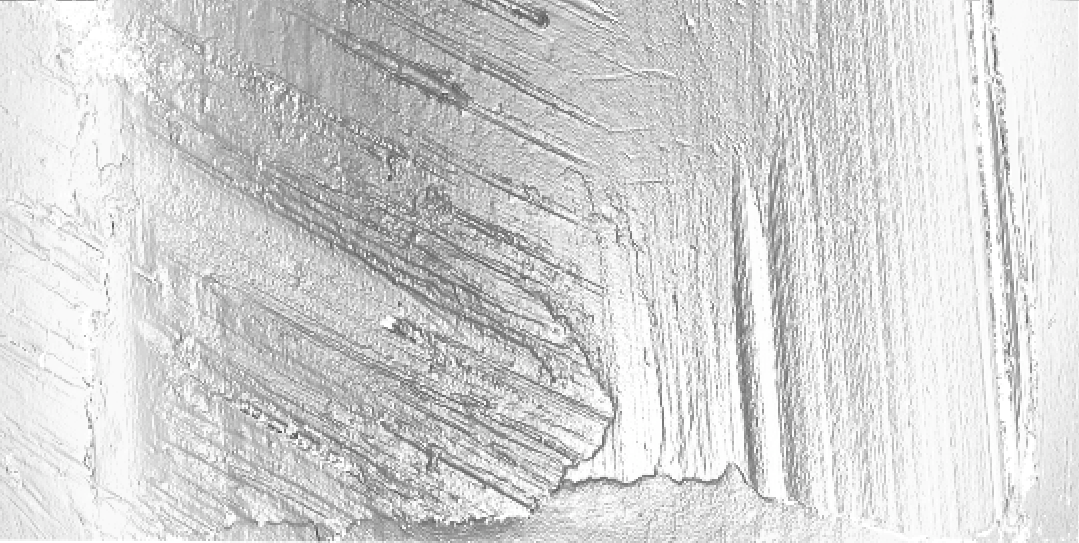}
\end{subfigure}
\begin{subfigure}[t]{.49\textwidth}\centering
\caption{Unknown Bullet B-2}
\includegraphics[width=\textwidth]{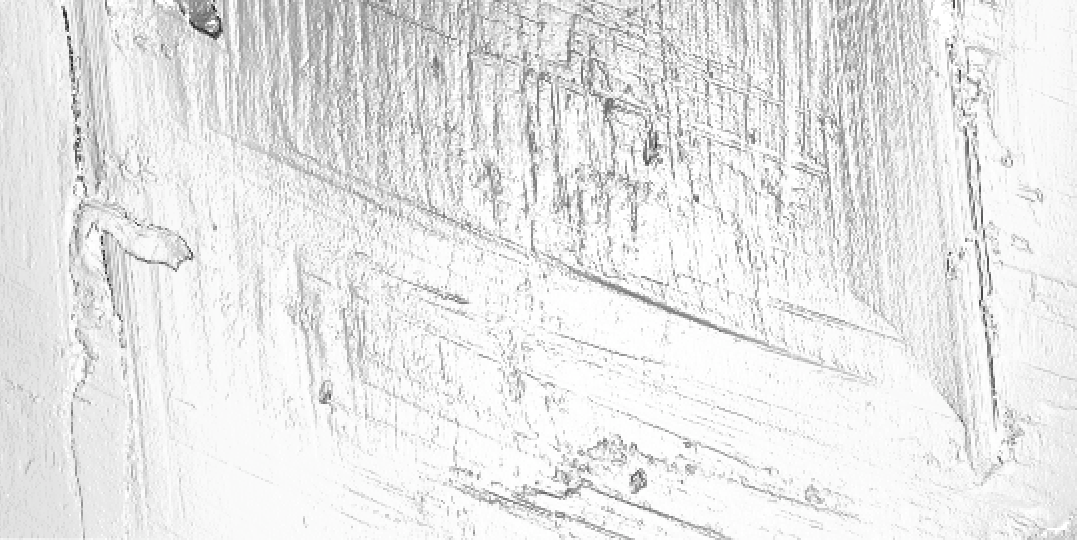}
\end{subfigure}
\begin{subfigure}[t]{.49\textwidth}\centering
\caption{Unknown Bullet Q-4}
\includegraphics[width=\textwidth]{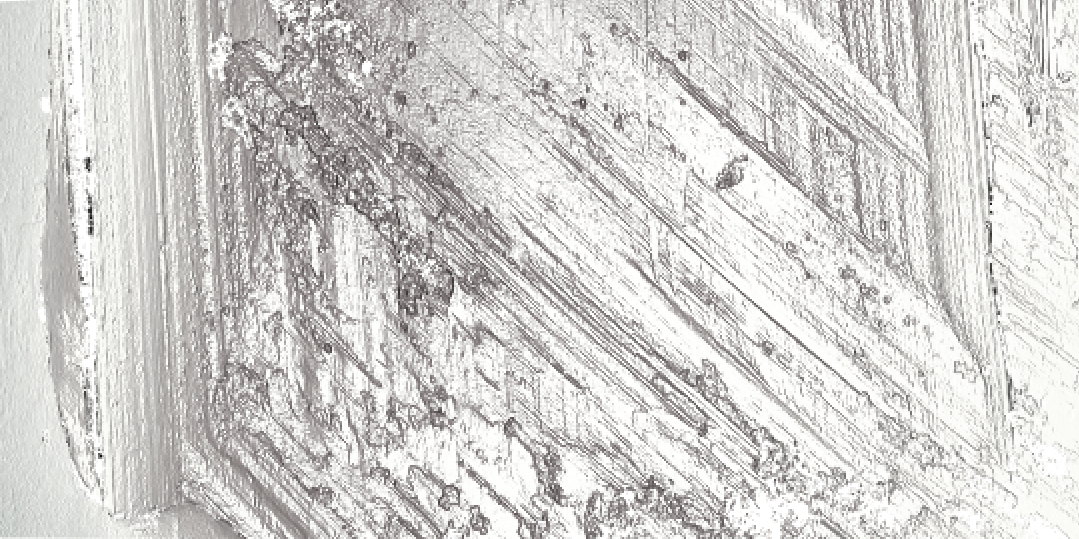}
\end{subfigure}
\caption{\label{fig:fourflags}Images of the four land impressions that got flagged during the quality assessment. All of them show scratch marks (tank rash) across the striation marks from the barrel. They are excluded from the remainder of the analysis.}
\end{figure}

\begin{figure}[hbtp]
\begin{subfigure}[t]{\textwidth}\centering
\caption{Maximal number of CMS between unknown bullet land impression C-3 and all of the other 118 considered (known) land impressions. For two land impressions the number of maximum CMS is high. }
\begin{knitrout}
\definecolor{shadecolor}{rgb}{0.969, 0.969, 0.969}\color{fgcolor}
\includegraphics[width=.5\textwidth]{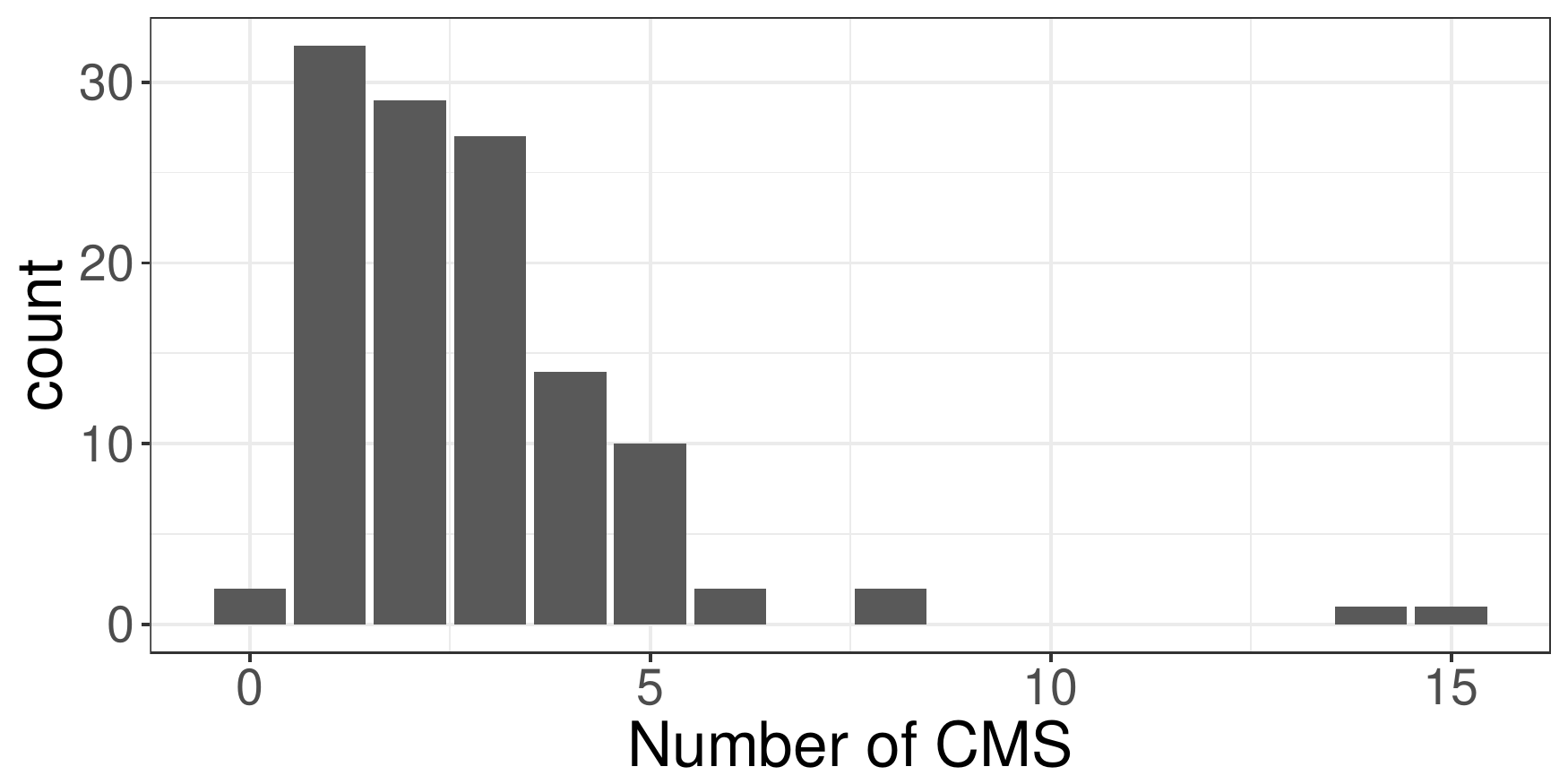}

\end{knitrout}
\end{subfigure}
\begin{subfigure}[b]{.49\textwidth}\centering
\caption{Overlaid signatures of C-3 and the land impression with the top matching CMS.}
\begin{knitrout}
\definecolor{shadecolor}{rgb}{0.969, 0.969, 0.969}\color{fgcolor}
\includegraphics[width=\textwidth]{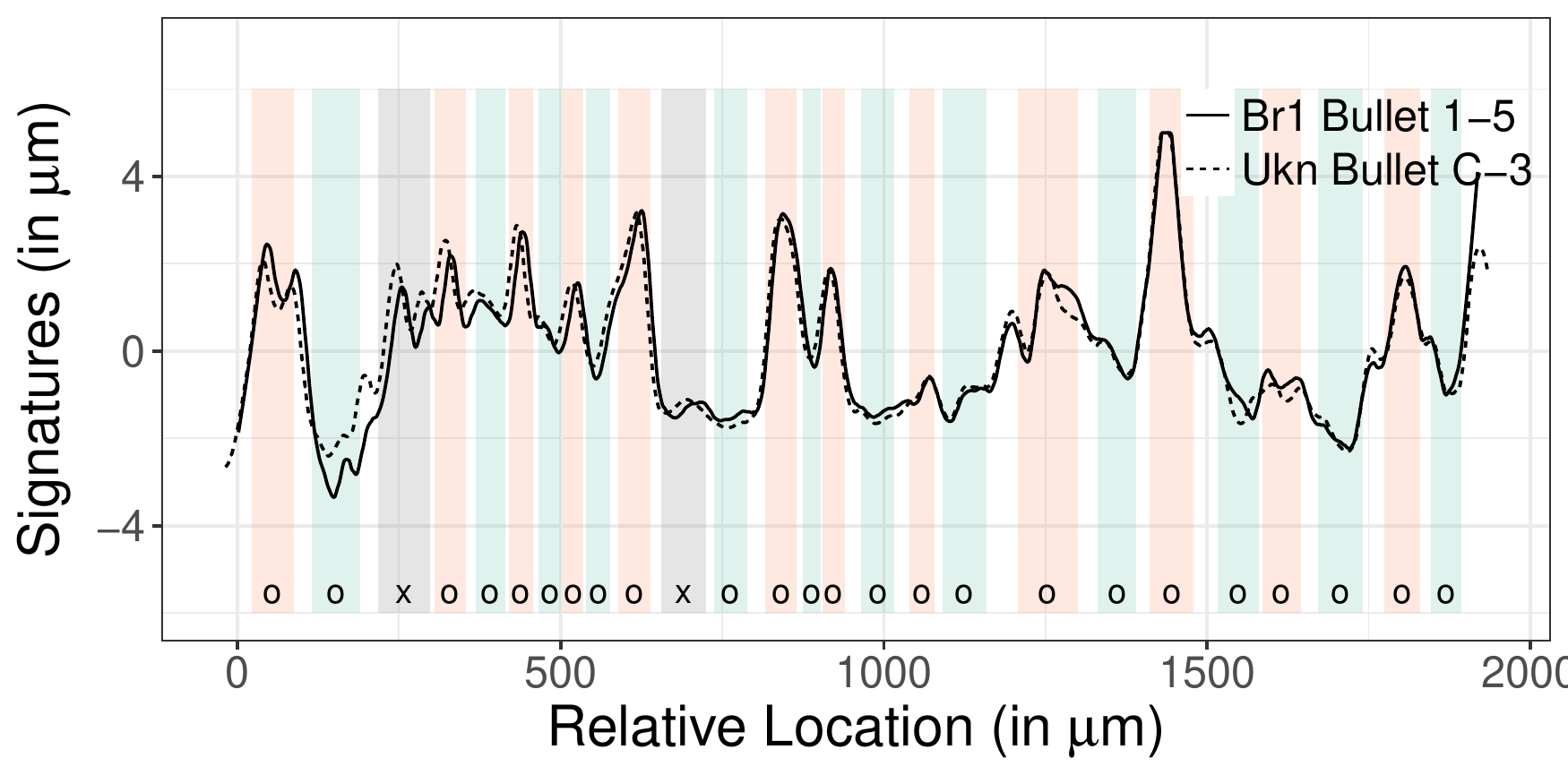}

\end{knitrout}
\end{subfigure}
\begin{subfigure}[b]{.49\textwidth}\centering
\caption{Top 2 match with C-3 based on CMS.}
\begin{knitrout}
\definecolor{shadecolor}{rgb}{0.969, 0.969, 0.969}\color{fgcolor}
\includegraphics[width=\textwidth]{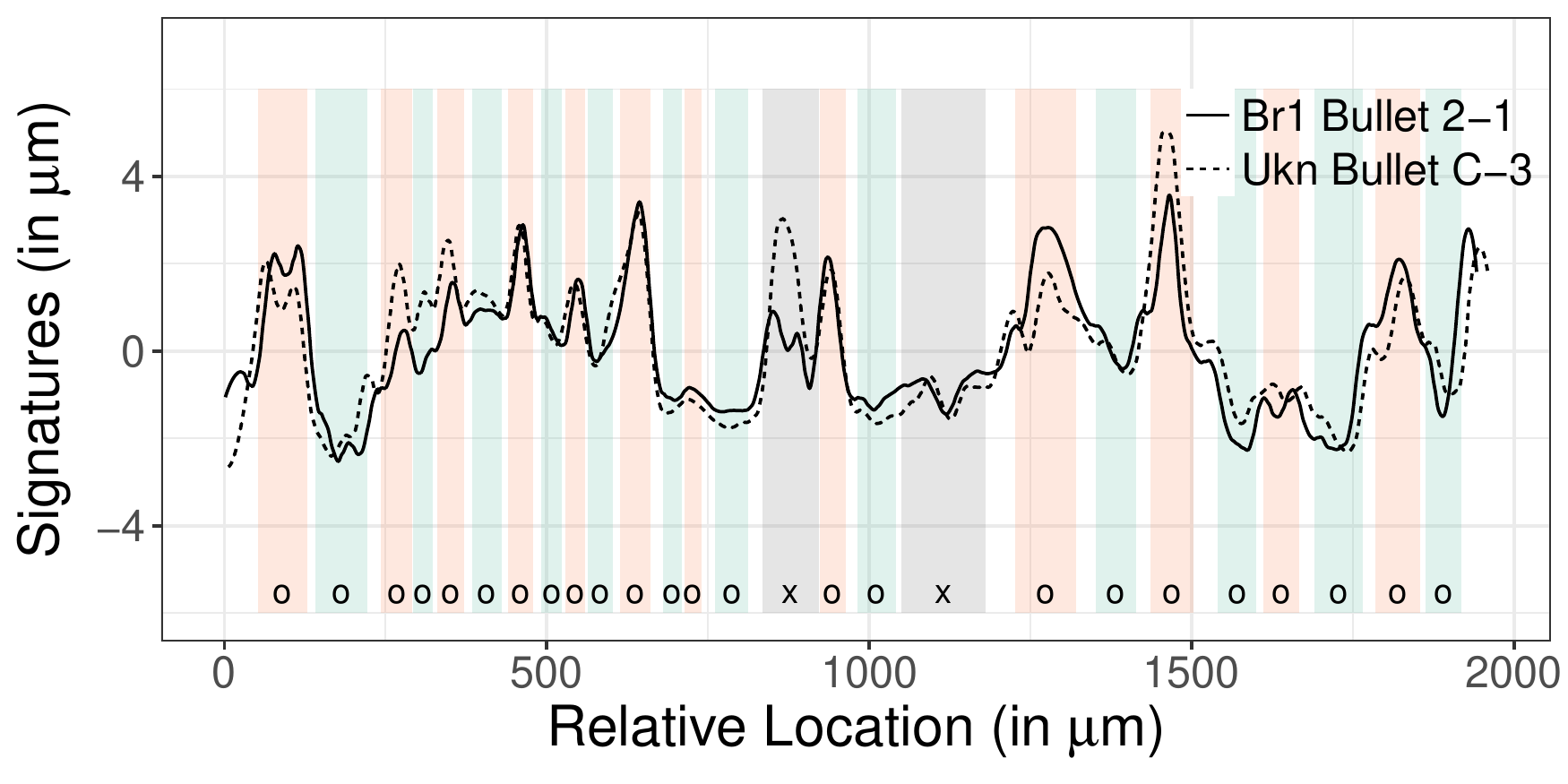}

\end{knitrout}
\end{subfigure}
\caption{\label{fig:hamby-perfect}Showcase scenario  when matching with CMS works very well. Unfortunately the matches are not always that convincing.}
\end{figure}

In order to get a better understanding of  how  the matching algorithm works in known matches and non-matches, we investigate its performance using the James Hamby study data. As a first step, we automatically assess the quality of each of the land impressions by checking that we can identify a stable region. For this, we compute the cross-correlation of signatures extracted from heights 25$\mu m$ apart. For a stable region, we require a minimum of 0.95 for the cross correlation. Four land impressions from different bullets are flagged as problematic in this respect. A visual inspection (see Figure~\ref{fig:fourflags}) shows that each one of these land impressions has scratch marks across the surface, also known as `tank rash' \citep{hamby:2009}.
We exclude these four land impressions from further matching considerations and
 run all remaining land impressions from the unknown bullets against all remaining land impressions from known bullets for matches, i.e. we are comparing $15 \times 6 -2 = 90 - 2 = 88$ land impressions from unknown bullets against $2 \times 10 \times 6 -2 = 120 - 2 = 118$ land impressions from known bullets, yielding a total of $10,384$ land-to-land comparisons. Out of these comparisons, there are 172 known matches (KM), while the rest are known non-matches (KNM).
Ideally, results look like the results in Figure~\ref{fig:hamby-perfect}: Figure~\ref{fig:hamby-perfect}a shows the distribution of the number of maximum consecutive matching striae between land impression C-3 and all 118 land impressions from known bullets. Two land impressions show a high CMS. These correspond to the known matches with C-3, shown in Figures~\ref{fig:hamby-perfect}b and~\ref{fig:hamby-perfect}c.
 Unfortunately, not all results are as clear cut.
It might not be reasonable to assume that we can match all land impressions, but the idea is to try to maximize the number of matches to get an overview of what we might be able to expect from an automated match.

\begin{figure}[hbtp]
  \centering
\begin{minipage}[t]{.47\textwidth}
\begin{knitrout}
\definecolor{shadecolor}{rgb}{0.969, 0.969, 0.969}\color{fgcolor}
\includegraphics[width=\textwidth]{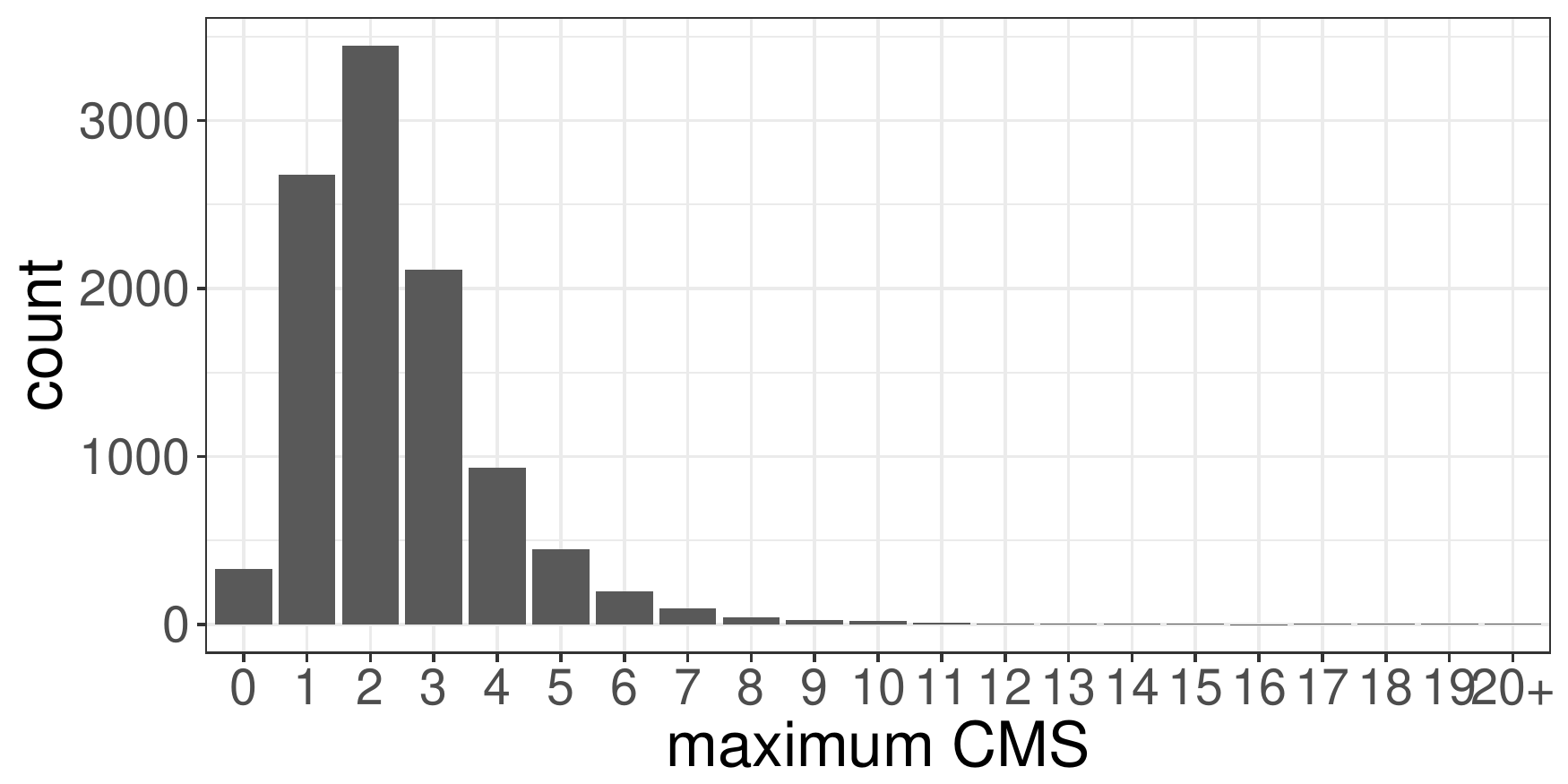}

\end{knitrout}
\end{minipage}
\begin{minipage}[t]{.52\textwidth}
\begin{knitrout}
\definecolor{shadecolor}{rgb}{0.969, 0.969, 0.969}\color{fgcolor}
\includegraphics[width=\textwidth]{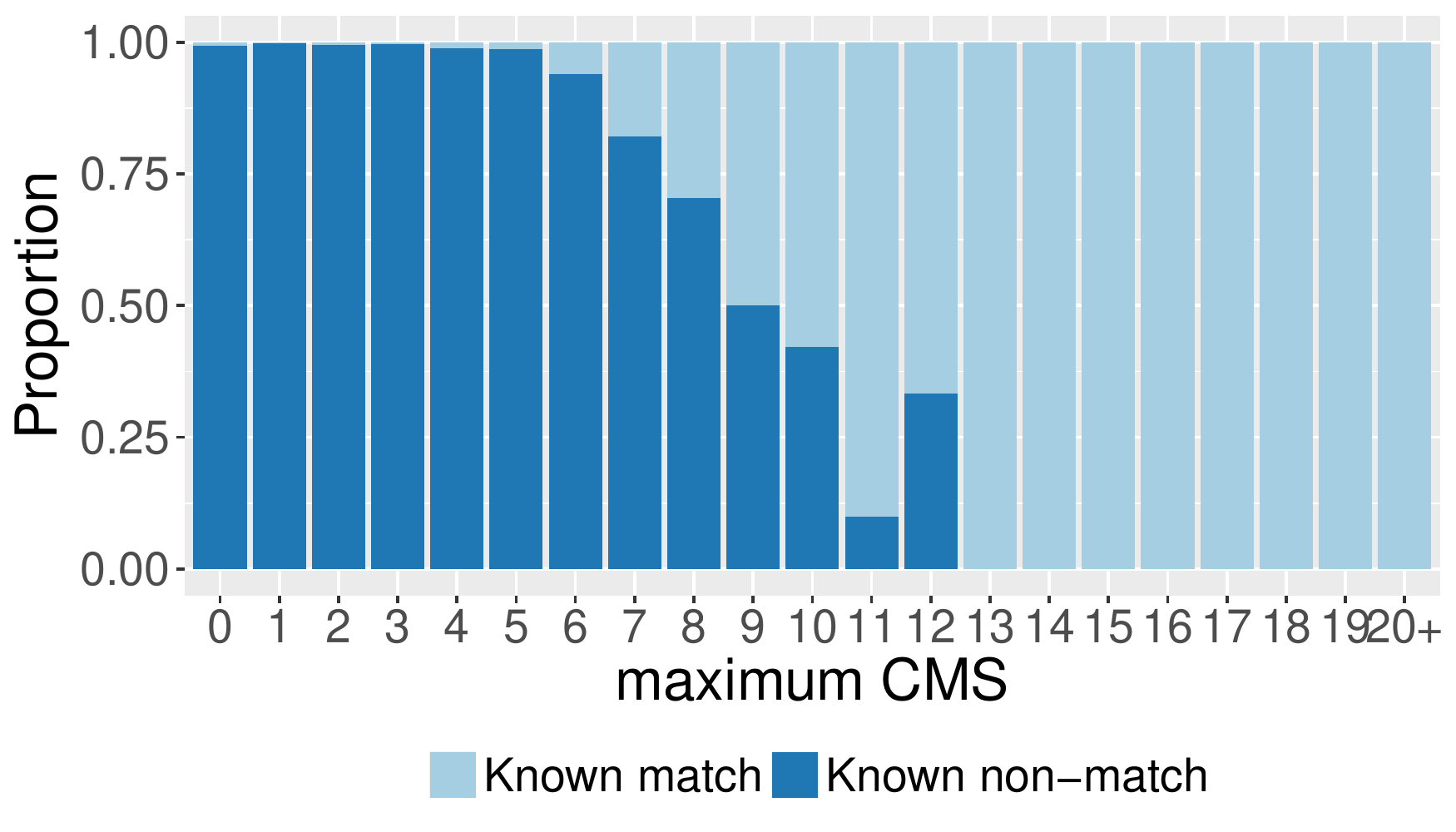}

\end{knitrout}
\end{minipage}
\caption{\label{fig:cms}Distribution of maximal CMS (left). Conditional barchart \citep{hummel} on the right: heights show probability of match/non-match given a specific CMS. All land-to-land comparisons with at least 13 CMS are matches.}
\end{figure}

\begin{figure}[hbtp]
  \centering
\begin{knitrout}
\definecolor{shadecolor}{rgb}{0.969, 0.969, 0.969}\color{fgcolor}
\includegraphics[width=.65\textwidth]{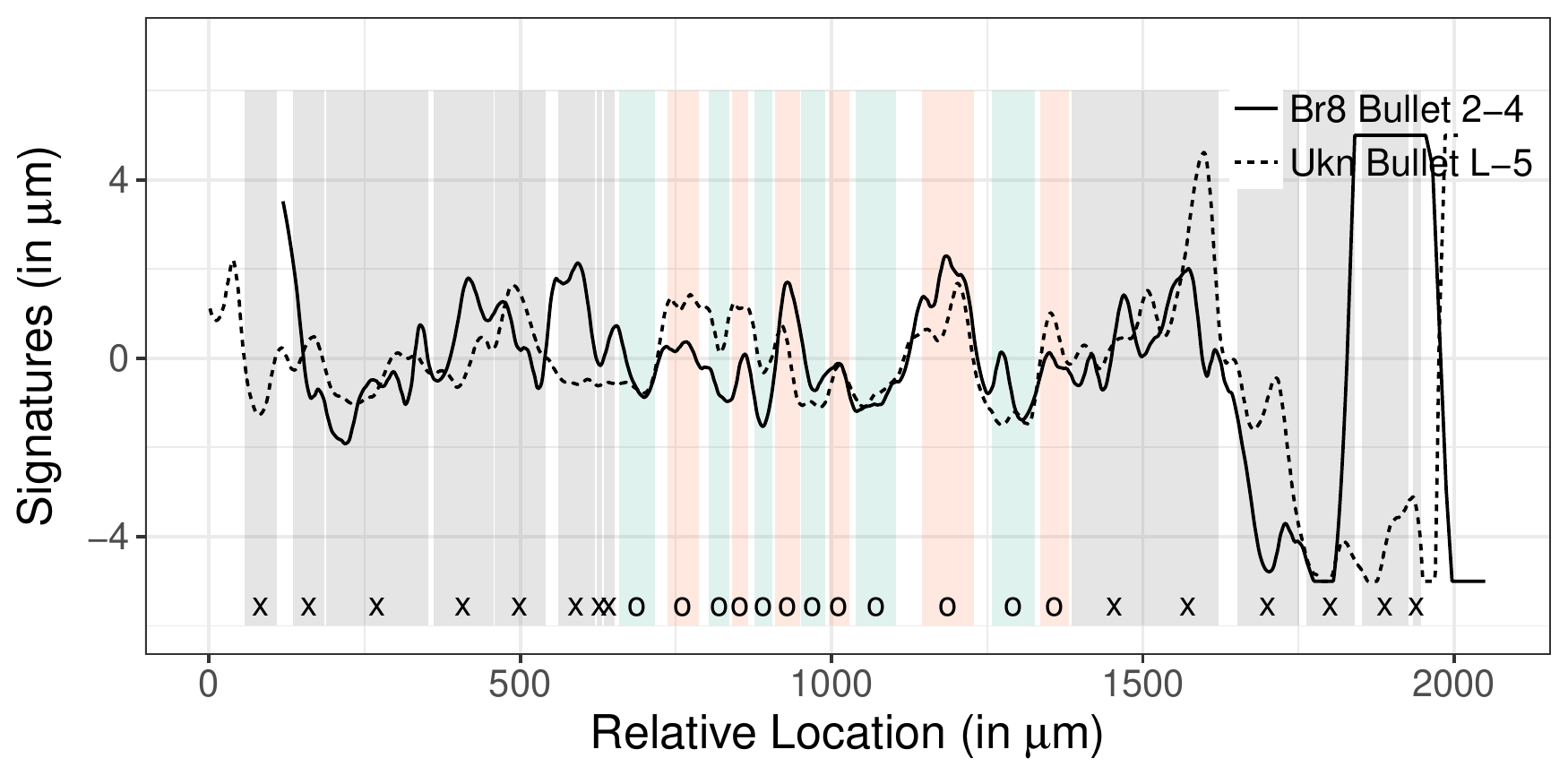}

\end{knitrout}
\caption{\label{fig:mismatch}Known mismatch with a relatively large number of maximal consecutive matching striae (twelve) in the middle. The pattern in the middle does look surprisingly similar, however the outer ends of the signatures easily reveals this comparison as mismatch. }
\end{figure}

Figure~\ref{fig:cms} shows the strong connection between the maximal number of consecutive striae and matches in the Hamby study. All 42 pairs of land impressions with at least thirteen CMS in common are matches.
There are two things that should be noted at this point: the automated algorithm finds a relatively high number of CMS even for non-matches. On average, there are 2.31 maximal CMS between known non-matches (with a standard deviation of 1.4). Known matches share on average 8.49 maximal CMS, with a standard deviation of 5.65. While the probability for a match increases with the number of maximal CMS, a large number of maximal CMS by itself is not indicative of a match, as was previously pointed out by \citet{miller:1998}. Figure~\ref{fig:mismatch} shows a known mismatch between two land impressions that share twelve consecutively matched striae. Visually we can easily tell that these two land impressions do not match well.

\begin{figure}[hbtp]
  \centering
\begin{knitrout}
\definecolor{shadecolor}{rgb}{0.969, 0.969, 0.969}\color{fgcolor}
\includegraphics[width=\textwidth]{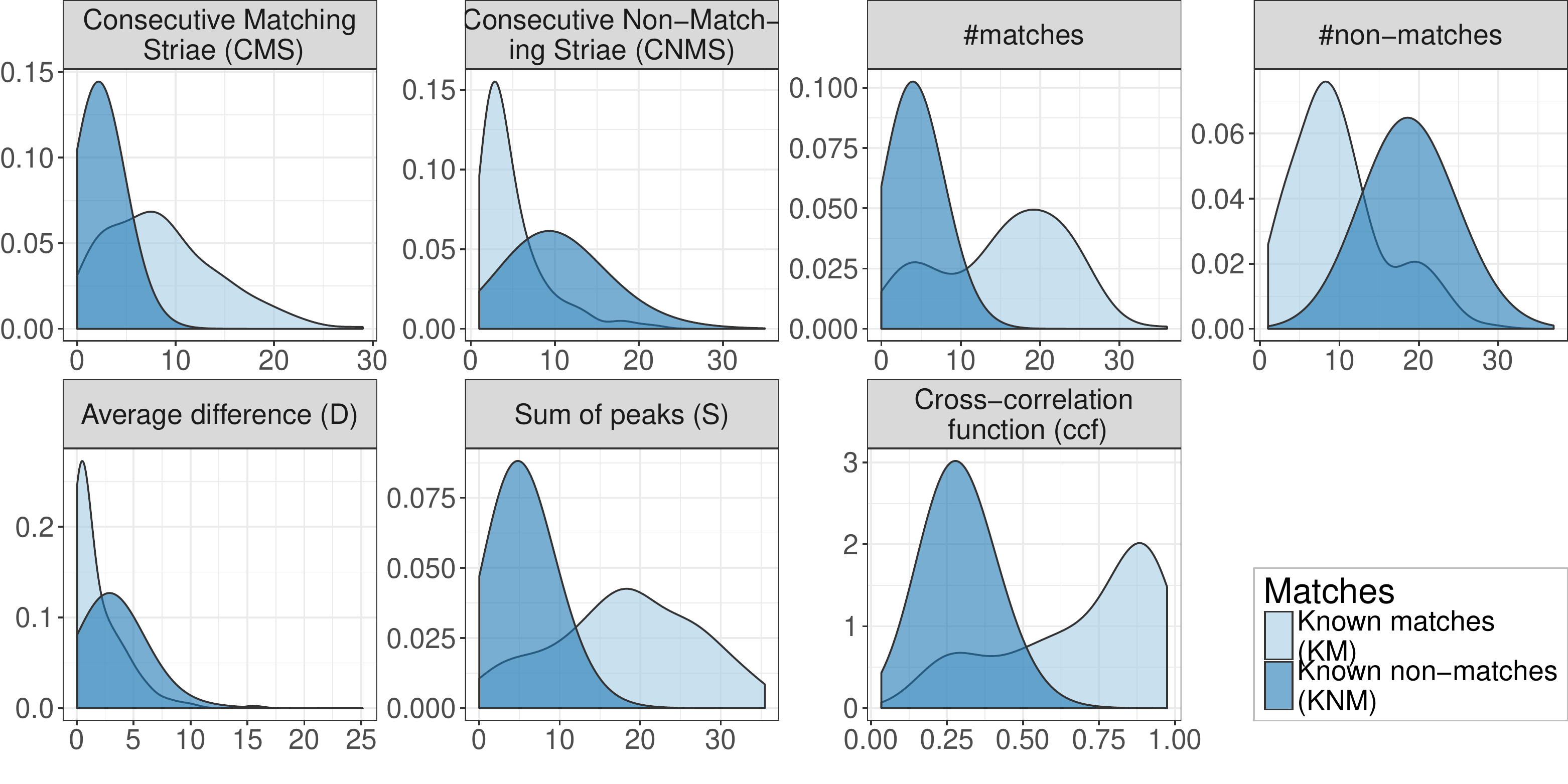}

\end{knitrout}
\caption{\label{fig:densities}Overview of all the marginal densities for features described in section~\ref{sec:algorithm}. Shifts in the mode of the density functions between known matches and known non-matches indicate the variable's predictive power in distinguishing matches and non-matches. Predictive power is shown in more detail in Figure~\ref{fig:rocs}.}
\end{figure}

For smaller numbers of CMS, the percentage of false positives quickly increases. However, if we take other features of the image into account, we can increase the number of correct matches considerably: Figure~\ref{fig:densities} gives an overview of the densities of all of the features derived earlier, for known matches (KM) and known non-matches (KNM). The densities of almost all of the features show strong differences between matches and non matches. For example, a high amount of cross-correlation between two signatures is indicative of a match --  in the Hamby study, only known matches have a cross-correlation of 0.75 or higher. There are 97 land-to-land comparisons with a cross-correlation that high.

\begin{figure}[hbtp]
  \centering
\begin{knitrout}
\definecolor{shadecolor}{rgb}{0.969, 0.969, 0.969}\color{fgcolor}
\includegraphics[width=\textwidth]{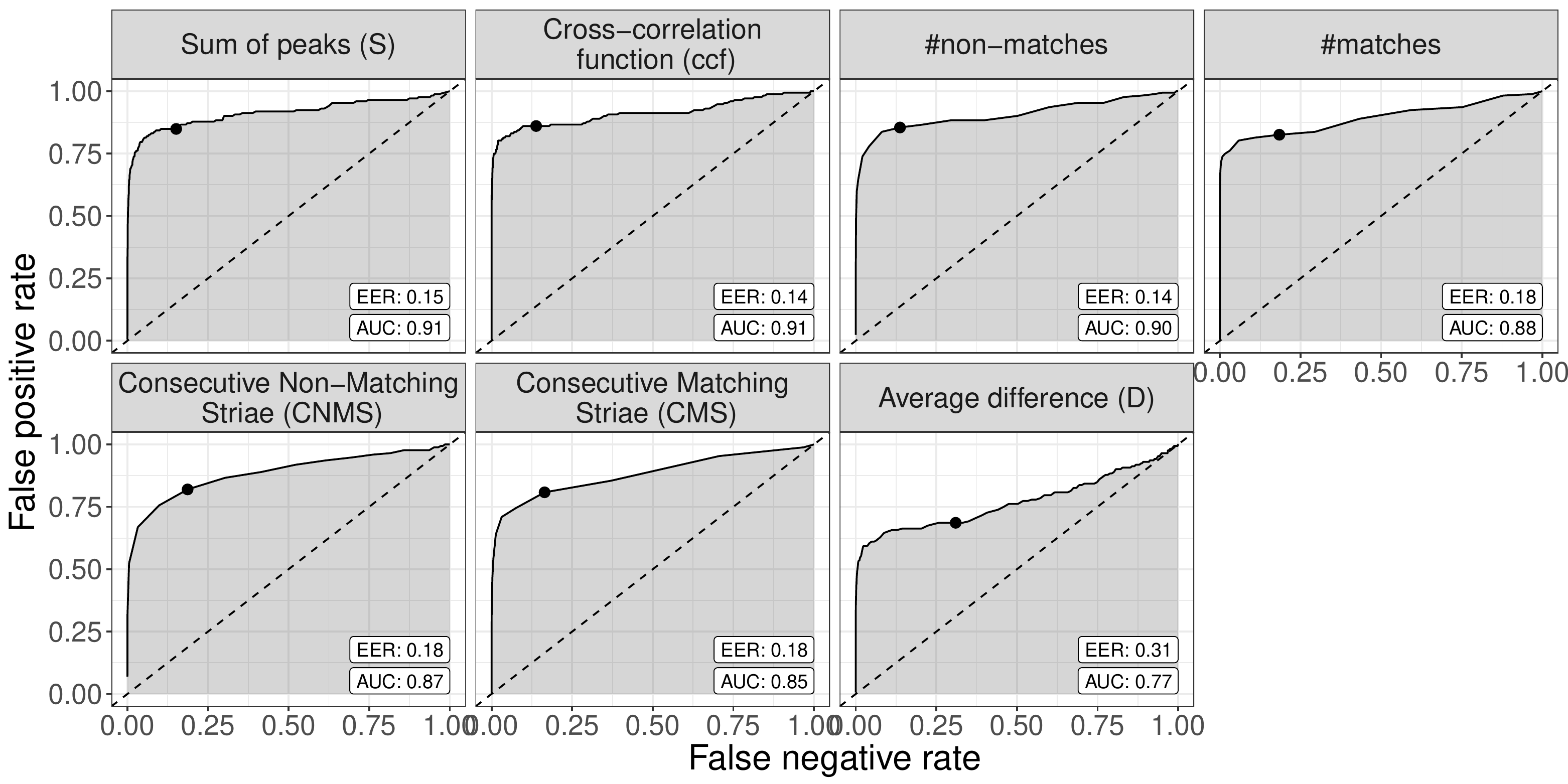}

\end{knitrout}
\caption{\label{fig:rocs}ROC curves for all of the features described in section~\ref{sec:algorithm}. Variables are sorted according to their area under the curve (AUC). The equal error rate (EER) is marked by a point on the ROC curve. Except for the distance $D$ between signatures, all individual features derived from the surface measurements and the aligned striation marks are more predictive than the maximal CMS. }
\end{figure}

All of the features in Figure~\ref{fig:densities} show large, if not significant, differences between matches and non-matches. The predictive power of each one of these features is shown in the form of the Receiving Operating Characteristic (ROC) curves in Figure~\ref{fig:rocs}. The features are arranged in descending order according to the area under the curve (AUC).
The dots mark the equal error rate, i.e. the location on the ROC curve, where false positive and false negative error rates are the same. The smaller the value, the better. We see that in this instance a low equal error rate (EER) goes hand in hand with high predictive power as measured in AUC.
The feature with the highest individual predictive power is $S$, the sum of the average heights of two signatures at peaks and valleys. The maximal number of CMS is only in the seventh position here. The overall high AUC values indicate that we can successfully employ machine learning methods to distinguish matches from non-matches.

\begin{figure}[hbtp]
  \centering
\begin{knitrout}
\definecolor{shadecolor}{rgb}{0.969, 0.969, 0.969}\color{fgcolor}
\includegraphics[width=.7\textwidth]{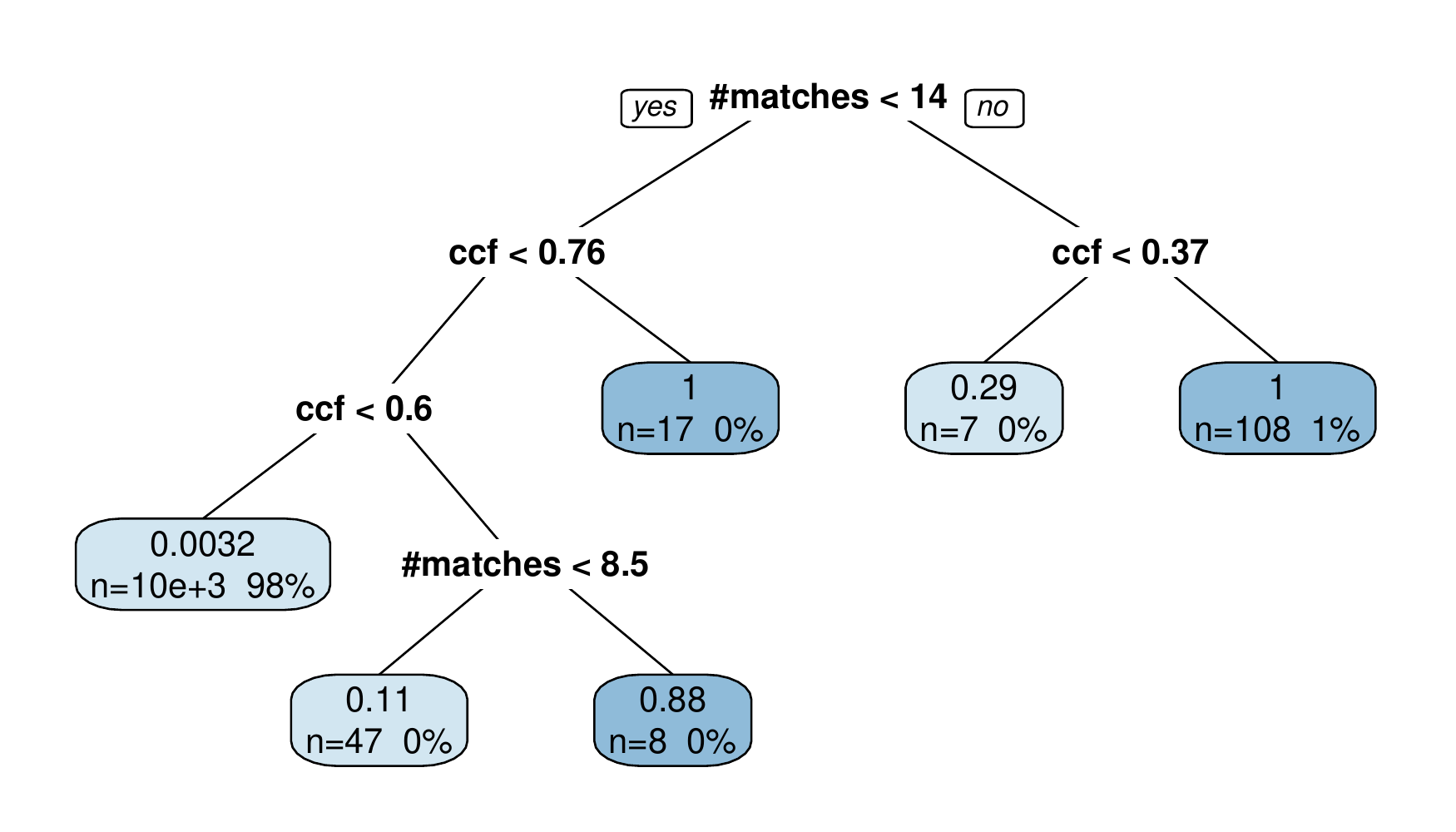}

\end{knitrout}
\caption{\label{fig:tree}Decision tree of matching bullets based on recursive partitioning. The rectangular nodes are the leaves, giving a short summary consisting of the number of observations in the leaf (bottom left), the corresponding percentage of the total (bottom right). The number at the top shows the fraction of these observations that are a match. A 1 or a 0 therefore indicate a homogeneous (or perfect) node. }
\end{figure}

Using recursive partitioning, we fit a decision tree~\citep{breiman:1984, rpart, rpart.plot} to predict matches between land impressions based on features derived from the image files. The resulting tree is shown in Figure~\ref{fig:tree}. A total of 132 land impressions is being matched correctly. Interestingly, the number of consecutive matching striae does not feature in this evaluation.
Instead of CMS, cross-correlation (ccf) between the signatures is very important in the matching process  by the decision tree. Aside from  cross correlation, the total number of matches is also included in the decision rule.
Between cross-correlation and CMS, cross-correlation has higher  predictive power. This  does not  contradict earlier findings emphasizing the value of CMS on visual assessments of bullet matches: in those papers, assessments were based on purely visual inspection of either actual bullets or 2D microscopic images of bullets.
Neither one of these methods allows for an assessment of cross-correlations. This is one of the benefits of switching to a digitized version of the images that preserves the 3D surface structure. The findings about the discriminating power of cross-correlation are consistent with the results of the study by \citet{ma:2004}. However, in that study, the authors did not consider the number of matches and non-matches.

\begin{figure}[hbtp]
  \centering
\begin{knitrout}
\definecolor{shadecolor}{rgb}{0.969, 0.969, 0.969}\color{fgcolor}
\includegraphics[width=.7\textwidth]{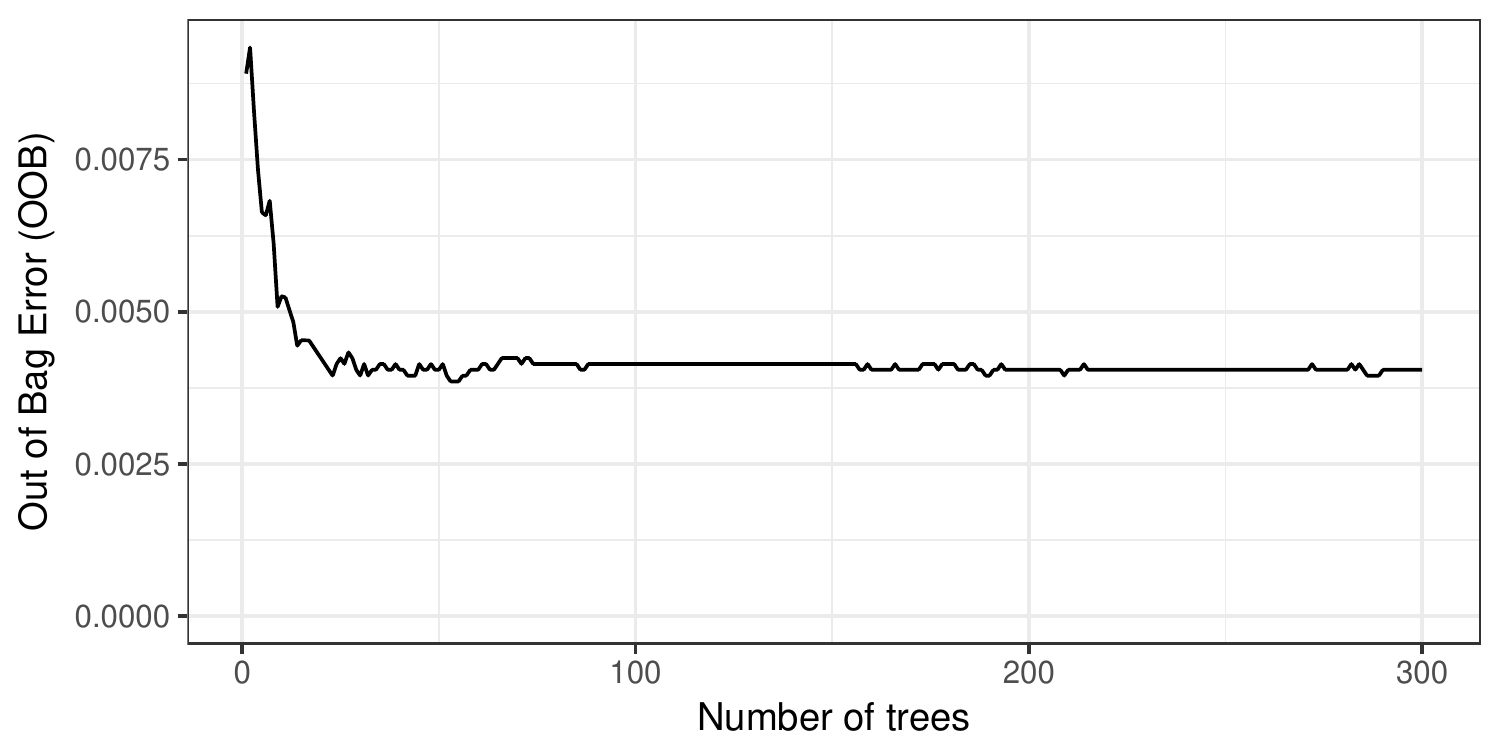}

\end{knitrout}
\caption{\label{fig:oob}Cumulative out-of-bag error rate of a random forest fit to predict land-to-land matches from image features.}
\end{figure}

\begin{figure}[hbtp]
  \centering
\begin{knitrout}
\definecolor{shadecolor}{rgb}{0.969, 0.969, 0.969}\color{fgcolor}
\includegraphics[width=.6\textwidth]{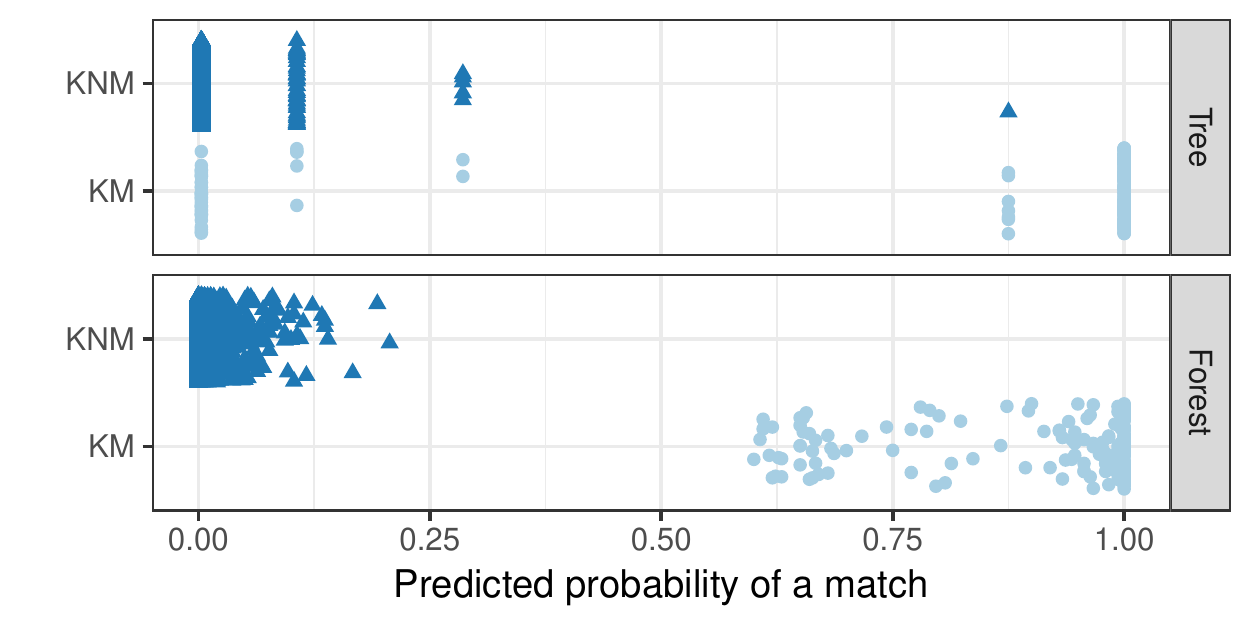}

\end{knitrout}
\caption{\label{fig:treeforest}Prediction results from the tree and the forest. Using a cut-off probability of 0.5 the forest correctly predicts every single comparison. Compared to the tree, the forest's prediction probabilities are  shrunk towards either end of the prediction range.}
\end{figure}

Another benefit of the digitized version of the images is that we can apply several hundred decision trees to combine in a random forest~\citep{breiman:2001, randomForest}.  For each of the trees in a random forest, only two thirds of the observations are used for fitting, while  the remaining third is used to evaluate the tree's predictive power and accuracy, or its reverse, the error rate. Because errors are determined from the one third of held-back observations, this error rate is called the out-of bag (OOB) error.
Figure~\ref{fig:oob} shows the cumulative out-of-bag error (OOB) rate for 300 trees.
After about 100 trees, the error rate of land-to-land comparisons stabilizes at 0.0039. This is a weighted average between false positive error rate of 0.0001 and an error rate of false negatives of 0.2267. This out-of-bag error rate is over-estimating the actual error in the Hamby study: here, the final random forest based on 300 trees is able to correctly predict all known  matches and non-matches (see Figure~\ref{fig:treeforest}).
Note that this error rate is based on land-to-land comparisons and is expected to be much lower for bullet-to-bullet comparisons. In the case of the Hamby data, even a single tree results in an overall error rate of zero, if we require that a match of two bullets occurs when at least two of the bullet's land impressions are matched. 
This makes the errors in the automated approach smaller than the human error in the Hamby study. Out of the 507 participants who returned results, eight (out of $15 \times 507 = 7,605$) bullets were not matched conclusively, corresponding to a rate of 0.0011.

For the Hamby data, error rates based on bullet-to-bullet matches do not carry a lot of weight because of the small size of the study: fifteen unknown bullets are successfully matched to two pairs of ten bullets. Matching bullets can only be tested realistically in a much bigger experiment.
Another thing to note about  the random forest's error rates is that they are based on probability cutoffs of 0.5, i.e.\ whenever the predicted probability of a match exceeds 0.5, a match is declared. Basing this decision on a threshold fixed at 0.5 may not be the best approach. In practice, examiners are allowed a third option of `inconclusive'. On a probability spectrum of outcomes we could therefore introduce an interval of `inconclusive' results in the middle of the spectrum -- which turns out to be unnecessary in the Hamby study, because, here, the results from the random forest are very clear cut. Figure~\ref{fig:treeforest} shows a comparison of the predicted probabilities of a match by the tree and the random forest. As expected, the random forest provides a more realistic estimate of the uncertainty in the classification. 

\begin{figure}
\begin{knitrout}
\definecolor{shadecolor}{rgb}{0.969, 0.969, 0.969}\color{fgcolor}
\includegraphics[width=.7\textwidth]{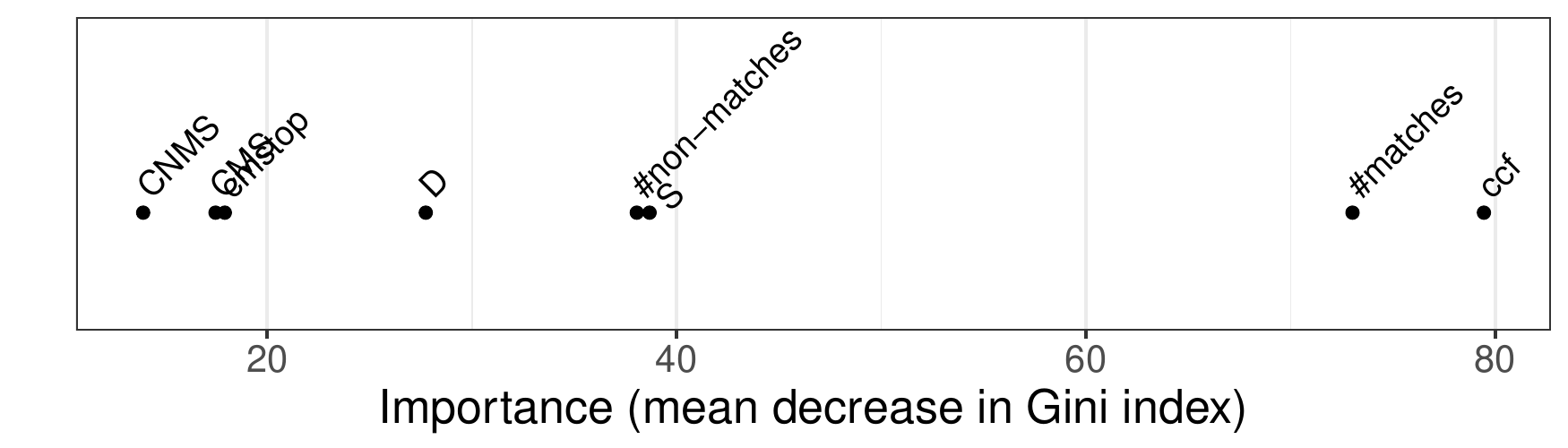}

\end{knitrout}
\caption{\label{fig:importance}Importance of features in the random forest. Importance is measured in terms of mean decrease in gini index when including the variable in a decision tree.}
\end{figure}

Besides resulting in a probabilistic quantification of matches, random forests also provide an assessment of the importance of each of the features derived from the bullets' 3D topological surface measurements. Figure~\ref{fig:importance} shows an overview of the importance of each variable measured as the mean decrease in the Gini index when the variable in question is included in a tree (for the exact values please refer to Supplement Section~\ref{supp:randomforest}).

The variables with the most predictive power are cross-correlation and the overall number of matching extrema, followed by the total depth of joint striations $S$ and total number of non-matches. CMS is found only in sixth place.

Besides including results from known matches against known non-matches, we can increase the number of comparisons in the Hamby study to include all possible land-to-land comparisons. This effectively doubles the number of data points available. Comparisons not previously included in fitting the random forest can also be used as an additional source for assessing error rates.  Results for this and a more detailed discussion can be found in Supplement Section~\ref{supp:extended}.

\section{Discussion}

We present an algorithm which detects the most prominent but least relevant structure of a bullet from a firearms identification perspective, removes these features, and produces residuals which allow for the easy identification of markings. We have generalized this algorithm to align the residuals from two bullets to automatically determine whether they are indistinguishable. A random forest model provides a probabilistic assessment of the strength of a match, along with an ordering of the relevance of features. Matching bullets is clearly not a one-step process, but rather a sequence of data analysis tasks each deserving attention. As there is no scientific standard in place at this point in time, our intent is to explain an approach to addressing these tasks, while documenting all steps and providing all code so other researchers and forensic scientists can reproduce and expand on our findings.

The matching algorithm is sensitive to the parameter choices made. The heights at which signatures are extracted (currently 25$\mu m$ apart) to evaluate stability, as well as the cross-correlation factor (currently 0.95) we set as a minimum threshold do affect the final outcome. Another parameter that must be selected is the amount of smoothing when identifying peaks and valleys (currently, a window of 23.4375$\mu m$ is used, corresponding to a window of 7 values to the left and the right of an observation). We try to lay out in the paper the impact that each of the parameter choices has on the matching performance, but more research and better data are needed to define an optimized scenario.

The Hamby study serves as our evaluation `database'. It consists of only  35 bullets -- this is obviously not a particularly realistic scenario for an automatic matching procedure, but for now we are unaware of other databases containing bullets in the x3p format that we could add to our study.

The feasibility of creating a database of images that could be used to identify guns used in crimes was evaluated in a 2008 report~\citep{nap:2008} by the National Research Council. The committee investigated the scalability of NIBIN (National Integrated Ballistic Information Network), which uses proprietary matching algorithms provided by IBIS. The bottom line of the report was that in spite of the many technical and practical hurdles, solutions to all but one problem could be found. The problem that remained is that statistically, the quality of the matching algorithm (in this case, of breech-face marks and firing pin impressions) could not withstand a hugely increased number of records without overwhelming forensic examiners, who have to examine possible matches suggested by the system.
The findings of the NRC report on imaging are based on two-dimensional greyscale images, which the committee argued were not reliable enough for distinguishing between fine marks. This finding coincides with the assessment by \citet{dekinder:2004} based on the IBIS Heritage system. A further re-assessment by \citet{deceuster:2015} came to the same conclusions based on the EvoFinder system.
The NRC report also found that results from 2D images can be improved when matches are based on 3D images. This is consistent with the importance of features found here: out of the top five features (see Figure~\ref{fig:importance}), only the total number of matches and mismatches are available for a match based on 2D features.

By suggesting an automated algorithm that first removes class characteristics, such as the groove impressions, shoulders, and the curvature of the bullet to reveal the region of the land impression, then identifies peaks and valleys on this land impression, we reduce subjectivity and with it possible sources of bias. In particular, `the concept of counting striations is subjective and based on experience'~\citep{miller:1998}. The steps outlined in this paper could also help explore other important forensic science problems. In particular, more general toolmark examination can benefit from the approach we discuss.

For a fair assessment of the performance of an algorithm, we need transparency. Our matching  algorithm is open: the code is readily available in form of the R package bulletr~\citep{bulletr}, and the code to produce this paper is available at \url{http://www.github.com/erichare/imaging-paper}. To understand whether an automated approach along the lines of the one we propose can accurately identify sets of bullets with undistinguishable markings, it will be necessary to assemble a much larger database that includes a wide range of ammunition types, degrees of damage, gun makes, etc. We are unaware of the existence of any such database. In addition to serving as a realistic testbed for the performance of the automated matching algorithm, such a database would also permit testing the underlying, as of yet untested, assumptions of uniqueness and reproducibility of the markings left by a gun on bullets.



\section*{Acknowledgment}
Thanks to David Baldwin for pointing us to the NIST database and doing a Firearms 101 for us.
Thanks to the men and women behind the software R~\citep{R}, and the authors of the R packages  knitr~\citep{knitr} and ggplot2~\citep{ggplot2}. We also wish to thank an anonymous reviewer, whose constructive and insightful comments contributed greatly to the improvement of the manuscript.


\bibliographystyle{imsart-nameyear.bst}
\bibliography{references}

\end{document}


\begin{frontmatter}

\title{Supplement to Automatic Matching of Bullet Land Impressions}
\runtitle{Supplement: Automatic Matching of Bullet Land Impressions}

\begin{aug}
\author{\fnms{Eric} \snm{Hare}\corref{}\ead[label=e1]{erichare@iastate.edu}\thanksref{m1}},
\author{\fnms{Heike} \snm{Hofmann}\ead[label=e2]{hofmann@iastate.edu}\thanksref{m1}}
\and
\author{\fnms{Alicia} \snm{Carriquiry}\ead[label=e3]{alicia@iastate.edu}\thanksref{m1}}

\affiliation{Iowa State University\thanksmark{m1}}

\runauthor{E. Hare et al.}
\end{aug}

\end{frontmatter}

\tableofcontents
\newpage

\section{Cylindrical Fit}\label{supp:cylindrical}
Figure~\ref{fig:fixedX} shows the  profile of surface measurements of bullet 1-5 at a fixed height. The smooth line on top is a circle, with estimated radius and center. The details of this fit are given below:
\begin{figure}[hbtp]
  \centering
\begin{knitrout}
\definecolor{shadecolor}{rgb}{0.969, 0.969, 0.969}\color{fgcolor}
\includegraphics[width=0.5\textwidth]{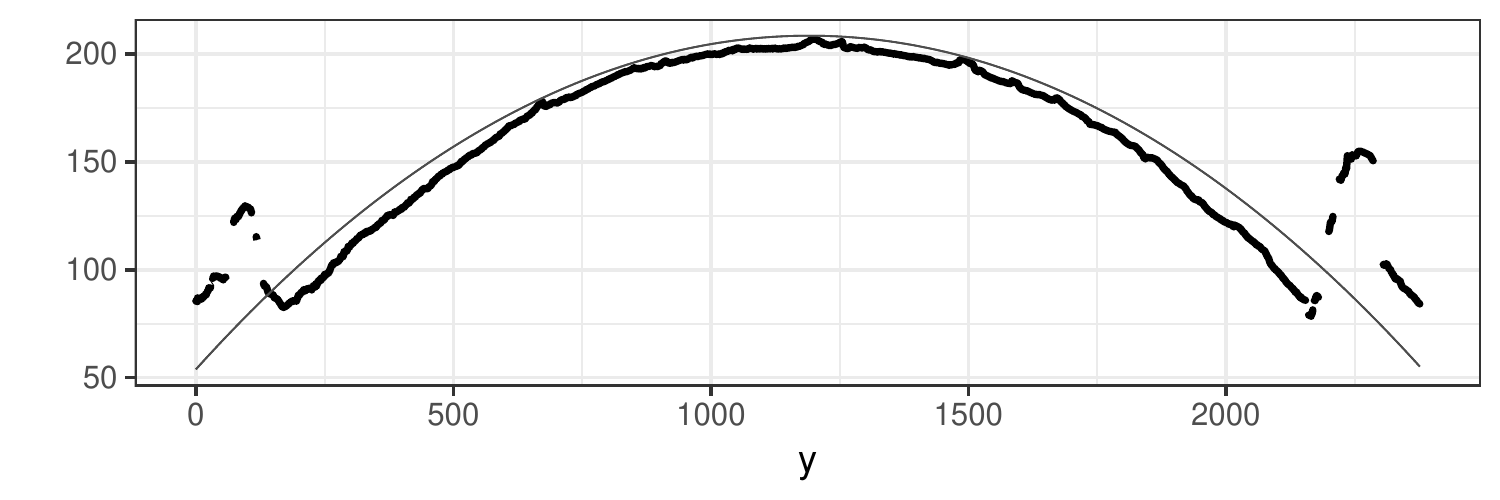}

\end{knitrout}
\caption{\label{fig:fixedX}Side profile of the surface measurements (in $\mu m$) of a bullet land at a fixed height of $x$. Note that the global features dominate any deviations, corresponding to the individual characteristics of striation marks.}
\end{figure}


Assume that $n$ data points are given in the form of data tuples $(x_1, y_1)$, $(x_2, y_2)$, $...$, $(x_n, y_n)$ that are (approximately) located on a circle. We want to estimate the location of the center and radius of the best fitting circle using a least squares approach.

We minimize the following expression:
\begin{equation}\label{eq:circle}
D = \sum_{i=1}^n \left( r^2 - (x_i-a)^2 - (y_i-b)^2 \right)^2,
\end{equation}
by differentiating $D$ with respect to $r, a,$ and $b$:
let us assume that $x_i$ and $y_i$ are centered (i.e. $\sum x_i = \sum_i y_i = 0$). Note, if they are not, make a note of the current means, subtract them now and add them to $(\hat{a}, \hat{b})$ at the end.

\noindent
The  derivate of $D$ with respect to $r$ is:
\begin{eqnarray*}
\frac{d}{dr} D &=& 2 \sum_i \left( r^2 - (x_i-a)^2 - (y_i-b)^2 \right) 2 r = \\
&=& 4 r \left( n r^2 - \sum_i (x_i-a)^2 - \sum_i(y_i-b)^2 \right).
\end{eqnarray*}
At the minimum:
\begin{equation}\label{eq:rmin}
\frac{d}{dr} D = 0 \stackrel{r \neq 0}{\iff} nr^2  = \sum_i (x_i-a)^2 + \sum_i(y_i-b)^2.
\end{equation}
The  derivative of $D$ with respect to $a$ is:
\begin{eqnarray*}
\frac{d}{da} D &=& 2 \sum_i \left( r^2 - (x_i-a)^2 - (y_i-b)^2 \right) 2 (x_i - a) = \\
&=& -4 \left[ a \cdot nr^2 + \sum_i (x_i - a)^3  + \sum_i (x_i - a) (y_i - b)^2 \right].
\end{eqnarray*}
Using (\ref{eq:rmin}) for $nr^2$  in the equation above we get:
\begin{eqnarray*}
\frac{d}{da} D &=& -4 \left[  \sum_i a(x_i-a)^2 +  \sum_i a(y_i-b)^2  + \right. \\
&& \phantom{-4 \ \ } \left . \sum_i (x_i - a)^3  + \sum_i (x_i - a) (y_i - b)^2 \right]  = \\
&=& -4 \left[ \sum_i (x_i-a)^2 (a + x_i - a)  + \right.\\
&& \phantom{-4 \ \ } \left .\sum_i (x_i - a + a) (y_i - b)^2 \right] = \\
&=& -4 \left[ \sum_i (x_i-a)^2 x_i   + \sum_i x_i  (y_i - b)^2 \right]
\stackrel{\begin{array}{c}\sum_i x_i = 0\\
\sum_i y_i = 0\end{array}}{=} \\
&=& -4 \left[ \sum_i x_i^3   + \sum_i x_i y_i^2  - 2a s_{xx} - 2b s_{xy} \right],
\end{eqnarray*}
where $s_{xx} = \sum_i x_i^2, s_{xy} = \sum_i x_i y_i$ and $s_{yy} = \sum_i y_i^2$.

\noindent
Likewise, we get for the derivative of $D$ with respect to $b$:
\begin{eqnarray*}
\frac{d}{db} D &=& -4 \left[ \sum_i y_i^3   + \sum_i x_i^2 y_i - 2a s_{xy} - 2b s_{yy} \right].
\end{eqnarray*}
To find the minimum we therefore get a system of two linear equations in $a$ and $b$:
\begin{eqnarray*}
2 s_{xx} a + 2 s_{xy} b = c_1 && \text{ with } c_1 = \sum_i x_i^3 + x_i y_i^2 \\
2 s_{xy} a + 2 s_{yy} b = c_2 &&\text{ with } c_2 = \sum_i x_i^2 y_i + y_i^3.
\end{eqnarray*}
The solution to the system is:
\begin{eqnarray*}
\hat{a} &=& \frac{c_1 s_{yy} - c_2 s_{xy}}{2 s_{xx} s_{yy} - 2 s_{xy}^2},\\
\hat{b} &=& \frac{c_2 s_{xx} - c_1 s_{xy}}{2 s_{xx} s_{yy} - 2 s_{xy}^2}, \text{ and}\\
\hat{r^2} &=& \frac{1}{n}s_{xx} + \frac{1}{n}s_{yy} + \hat{a}^2 + \hat{b}^2.
\end{eqnarray*}

\begin{figure}[hbtp]
  \centering
\begin{subfigure}[b]{.49\textwidth}\centering
\caption{\label{fig:residuala}Residual structure at height $x = 1.5625\mu m$ (bottom of the bullet).}
\begin{knitrout}
\definecolor{shadecolor}{rgb}{0.969, 0.969, 0.969}\color{fgcolor}
\includegraphics[width=\textwidth]{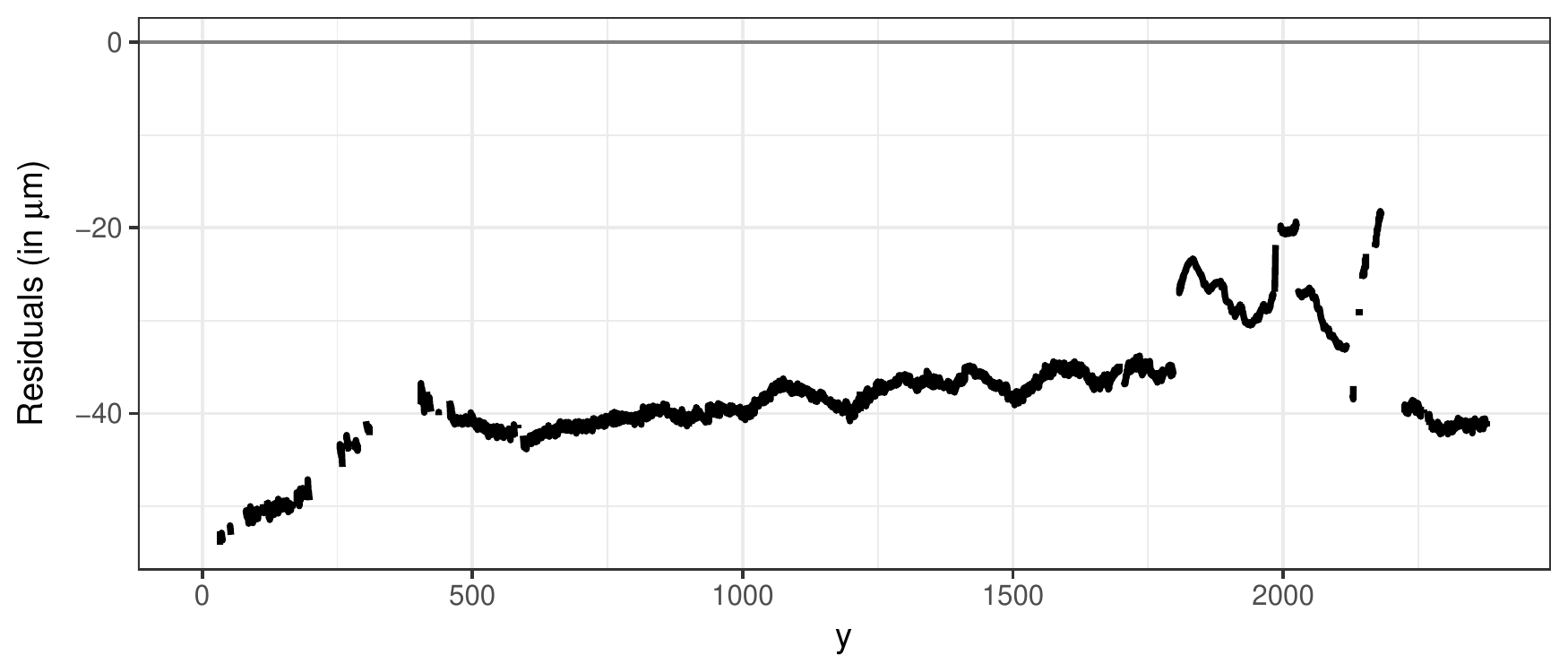}

\end{knitrout}
\end{subfigure}
\begin{subfigure}[b]{.49\textwidth}\centering
\caption{\label{fig:residualb} Residual structure at height $x = 100.00\mu m$}
\begin{knitrout}
\definecolor{shadecolor}{rgb}{0.969, 0.969, 0.969}\color{fgcolor}
\includegraphics[width=\textwidth]{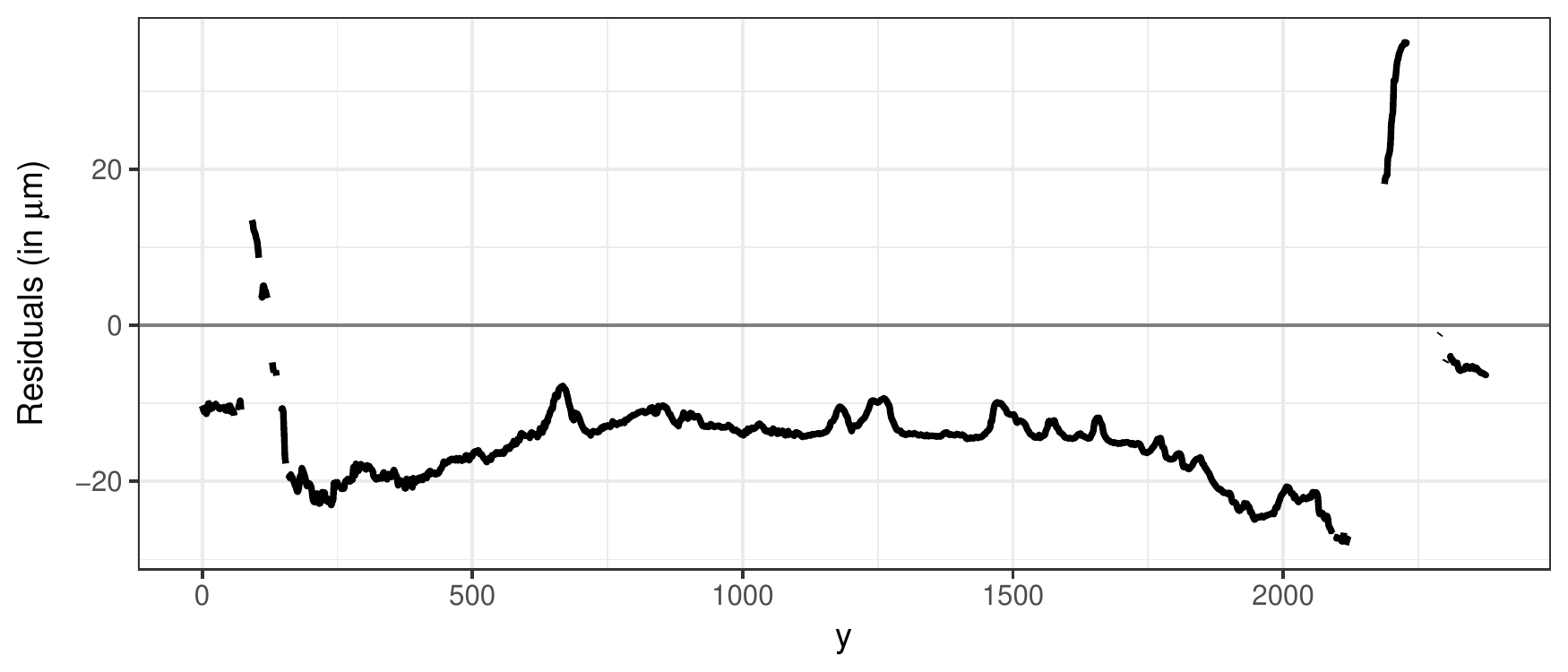}

\end{knitrout}
\end{subfigure}
\caption{\label{fig:residual} Residual structure of circular fits at two different cross sections. Both residual plots show systematic structures, indicating that a circular fit is not entirely appropriate.}
\end{figure}

The scatterplot in Figure~\ref{fig:residual} shows the residuals of such a fit.
In this instance, the radius is estimated as $\hat{r} = 4666.49\mu m = 4.67mm$ and the land impression covers about 29.5 degrees.  Both of these estimates are consistent with a 9 mm bullet fired by a Ruger P-85.
The residuals are dominated, as expected, by the shoulders, which show up as large positive residuals. For a profile at height $x = 100\mu m$ there is a residual circular structure that does not show up for all signatures.

\begin{figure}[hbtp]
  \centering
\begin{knitrout}
\definecolor{shadecolor}{rgb}{0.969, 0.969, 0.969}\color{fgcolor}
\includegraphics[width=\textwidth]{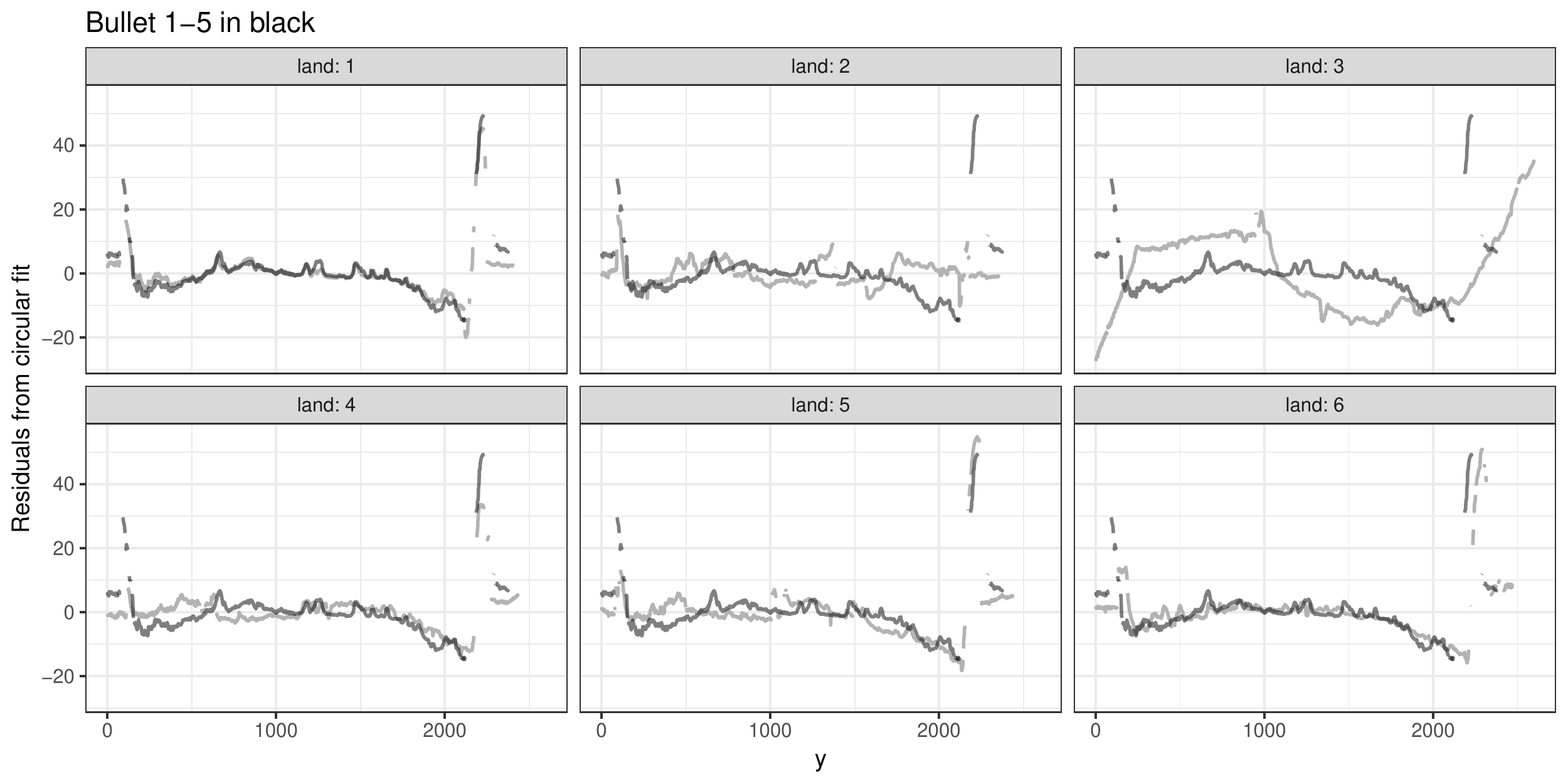}

\end{knitrout}
\caption{Circular fit to the signature of each land impression of bullet 2, with signature from bullet 1-5 overlaid.\label{fig:circlefits} The signature of bullet 1-5 matches best with bullet 2-1.}
\end{figure}

A single cylinder as a fit is unlikely to be a particularly good fit, because there seem to be quite massive deformations in the vertical direction. Even when we fit a circle at each distinct height of the bullet, as in Figure~\ref{fig:circlefits}, this does not address all of these issues. While the wider circumference at the base of the bullet can be resolved by individual circular fits, the systematic residual structure in Figure~\ref{fig:residualb} stays the same.

\newpage
\section{Assessing cross-correlation between signatures at multiple levels of height}\label{supp:ccf}

\begin{figure}[hbtp]
  \centering
\begin{knitrout}
\definecolor{shadecolor}{rgb}{0.969, 0.969, 0.969}\color{fgcolor}
\includegraphics[width=\textwidth]{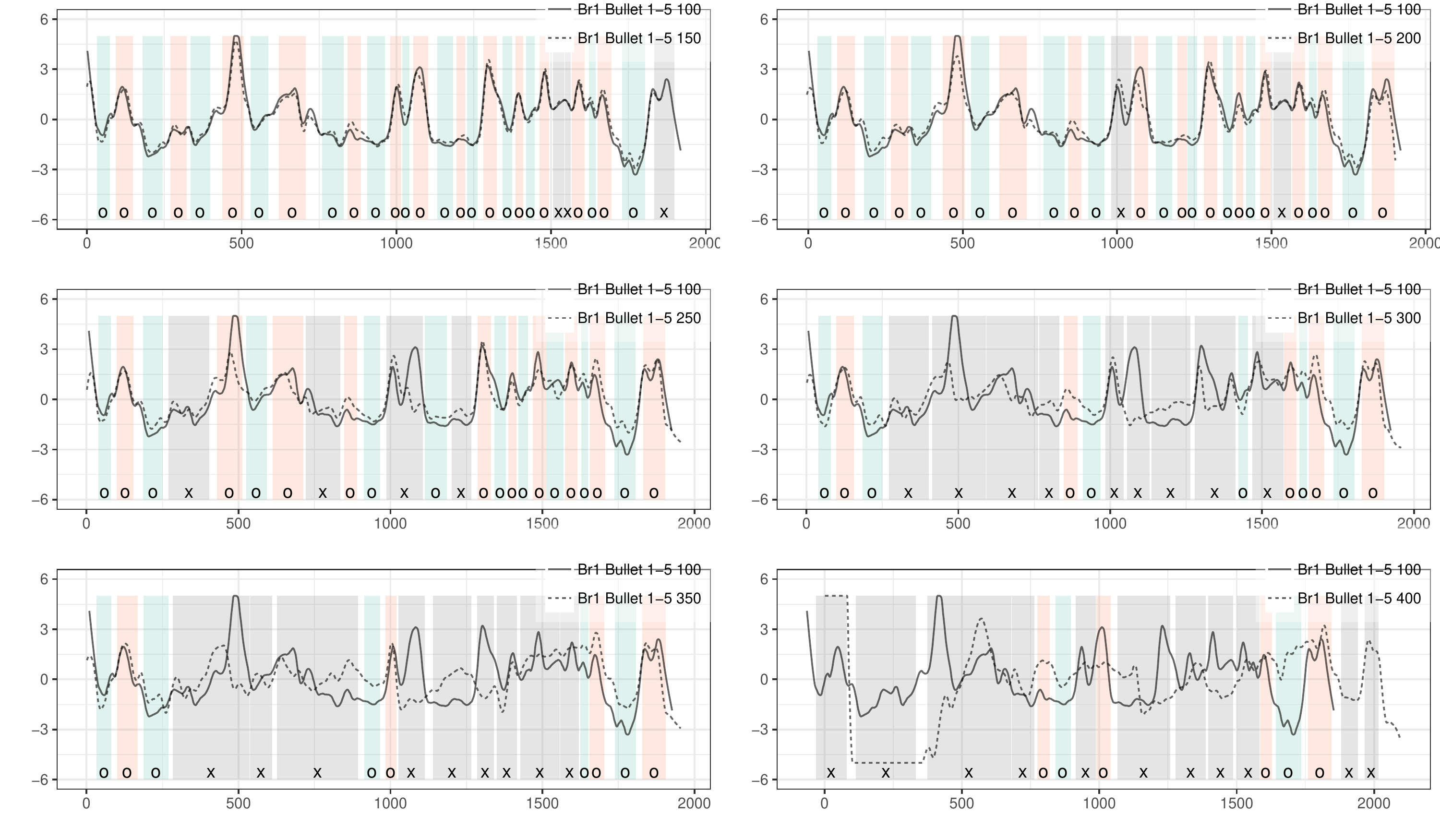}

\end{knitrout}
\caption{\label{fig:crosscuts}Overview of the variations in the signatures at different heights. The signature extracted at $x = 100\mu m$ is compared to signatures at every 50$\mu m$. With every step away from the original height, the number of differences between the signatures increases; the number of maximum CMS decreases from initially 22 to  four or fewer at a height of $x = 300\mu m$ and above. }
\end{figure}

Figure~\ref{fig:crosscuts} shows a sequence of signatures for bullet 1-5 (barrel 1) that are taken at heights 50$\mu m$ apart, between 150$\mu m$  and 400$\mu m$. These are compared to the signature at a height of 100$\mu m$. Initially, this comparison constitutes an almost perfect match between the two signatures. However, the match quickly deteriorates with increasing distance between the heights at which signatures are extracted.  Only if signatures are from heights  within 150$\mu m$ do we get a good visual match even when we know that the same bullet surface is being used.
Given that we have to expect some variation in nominally the same height values due to (manual) alignments in microscopes, we should take height values into account in the automatic matching routine by evaluating matches at several heights.

\newpage
\section{Signature intensities}\label{supp:bulletbottom}

\begin{figure}[hbtp]
\begin{knitrout}
\definecolor{shadecolor}{rgb}{0.969, 0.969, 0.969}\color{fgcolor}
\includegraphics[width=0.7\textwidth]{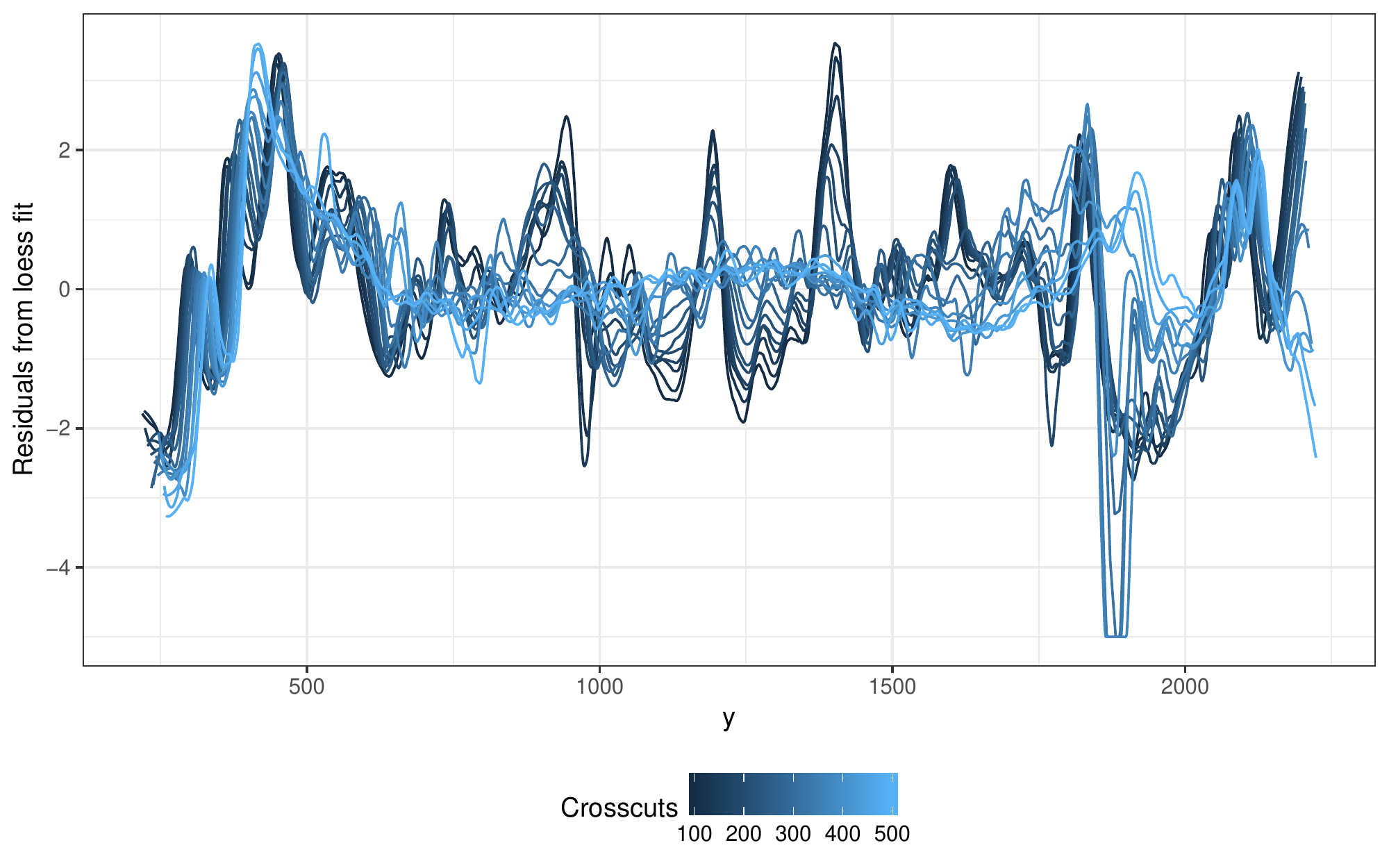}

\end{knitrout}
\caption{\label{fig:overview}Signatures of the same bullet at different heights.  With increasing height, peaks and valleys are less pronounced, resulting in a smaller standard deviation.}
\end{figure}

\begin{figure}[hbtp]
\centering
\begin{knitrout}
\definecolor{shadecolor}{rgb}{0.969, 0.969, 0.969}\color{fgcolor}
\includegraphics[width=.7\textwidth]{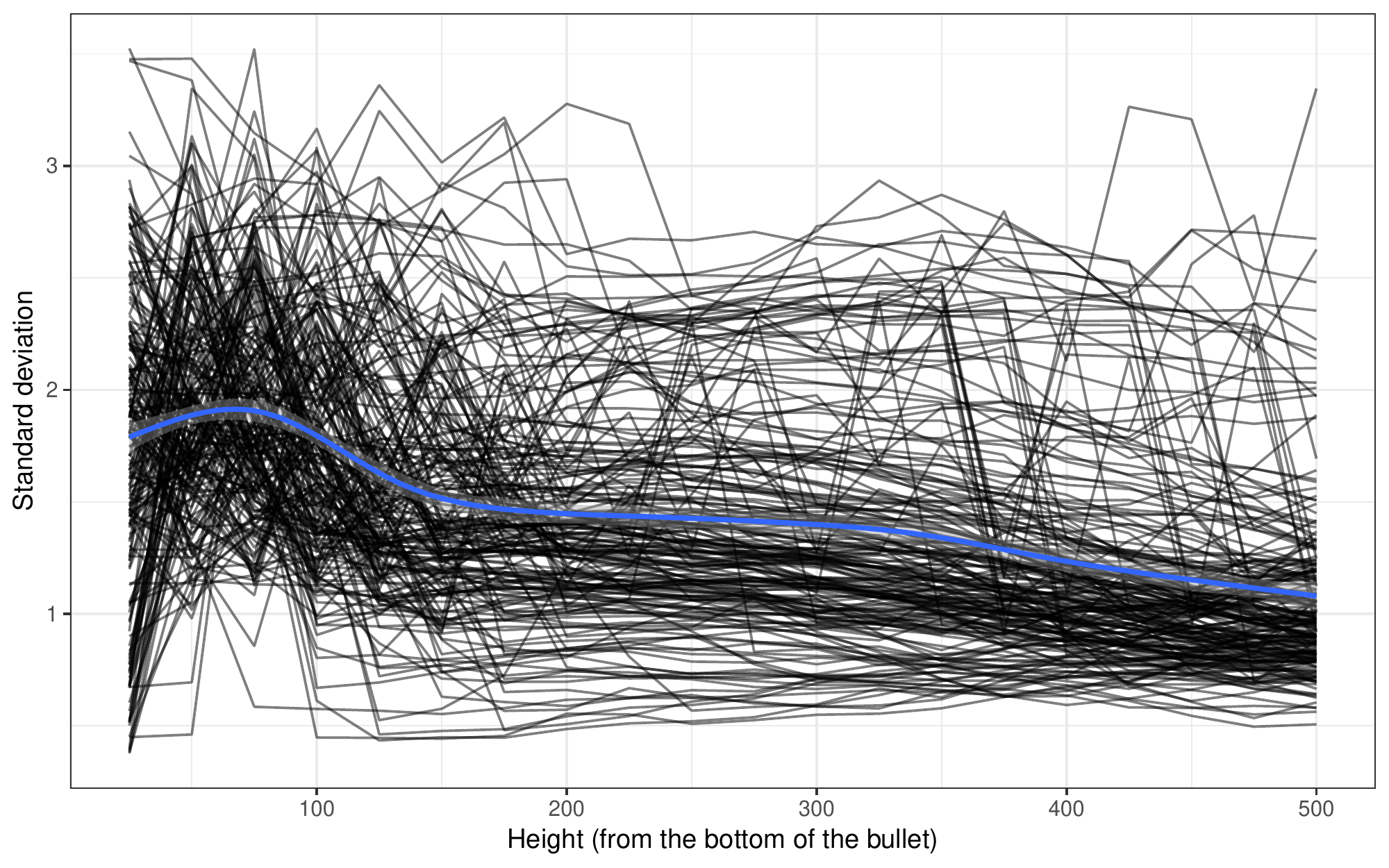}

\end{knitrout}
\caption{\label{fig:sds}Standard deviation reduces as height increase.}
\end{figure}

Figure~\ref{fig:overview} shows an overview of the signatures at different heights on a single bullet.
At larger heights  individual characteristics become less distinctive, making true matches to other bullets harder. The pattern of decreasing peaks and valleys is generally true for bullet land impressions, as can be seen in Figure~\ref{fig:sds}.
Figure~\ref{fig:sds} shows that the amount of standard deviation of a signature decreases on average for all bullet land impressions at larger heights.
This makes standard deviation of a signature one measure to quantify the extent to which a signature is expressed. For identifying matches we should therefore use the lowest height to extract a bullet's signature once a stable surface region is detected. This is in accordance with current standard practice \citep{afte:1992}.

\section{Complete evaluation of the Hamby study}\label{supp:extended}
One way to expand the use of the James Hamby study is to not only match all of the unknown bullet land impressions against the known bullet land impressions, but to compare every land impression against every other land impression. This effectively doubles the number of comparisons from 10,384 pairwise comparisons of usable bullet land impressions to 21,115 $\left[= (118+88)\cdot 205/2\right]$ comparisons by adding another 10,731 bullet land comparisons made up of known-to-known and unknown-to-unkown comparisons.

When we predict the new 10,731 comparisons using the random forest based on the previously considered 10,384 known-unknown comparisons, we encounter 18 false negatives and 9 false positives, corresponding to an actual false error rate of 0.19 and a false positive rate of 0.00085, which is close to the random forest's estimated OOB error rates of 0.226744 and 0.000098.

However, if we use all of the available comparisons to fit another random forest of 300 trees, the defacto error rates for false positives and false negatives are again at 0. The estimated OOB error rates are 0.00024 for the false positive rate and 0.22180 for the false negative rate. The false positive rate is therefore virtually unchanged, while we see a slight improvement in the false negative rate for an overall OOB error rate of 0.3\%, i.e.\ an increase to twice the number of comparisons leads to a decrease of 25\% of the estimated error rate. This is yet another argument in favor of a larger database for training algorithms.




Figures~\ref{fig:aligned} and~\ref{fig:aligned-second} give an overview of all the signatures from bullet land impressions in the Hamby study aligned by barrel. Three to five bullets were fired from each barrel. The figures give us both some insight into how well signatures match and how consistent individual characteristics are impregnated on bullets fired from each of the barrels. Signatures for some land impressions match remarkably well -- such as land 5 from barrel 1, whereas all land impressions from barrel 5 show some variability both in the location and depths of peaks and valleys.

\begin{landscape}
\begin{figure}[hbtp]
\begin{knitrout}
\definecolor{shadecolor}{rgb}{0.969, 0.969, 0.969}\color{fgcolor}
\includegraphics[width=7.5in]{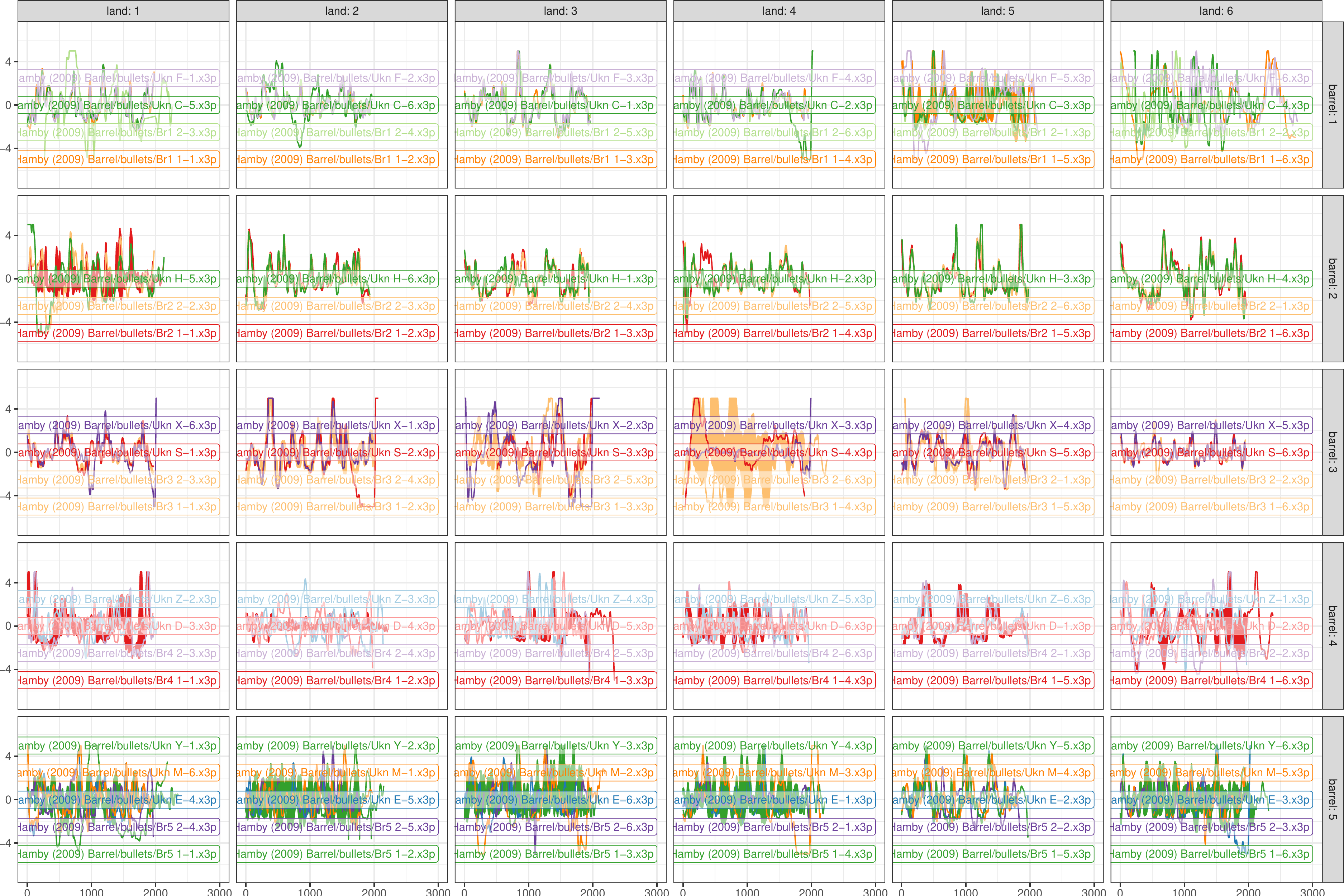}

\end{knitrout}
\caption{\label{fig:aligned}Overview of aligned signatures for all bullet land impressions for  barrels 1 to 5 of the Hamby study. }
\end{figure}

\begin{figure}[hbtp]
\begin{knitrout}
\definecolor{shadecolor}{rgb}{0.969, 0.969, 0.969}\color{fgcolor}
\includegraphics[width=7.5in]{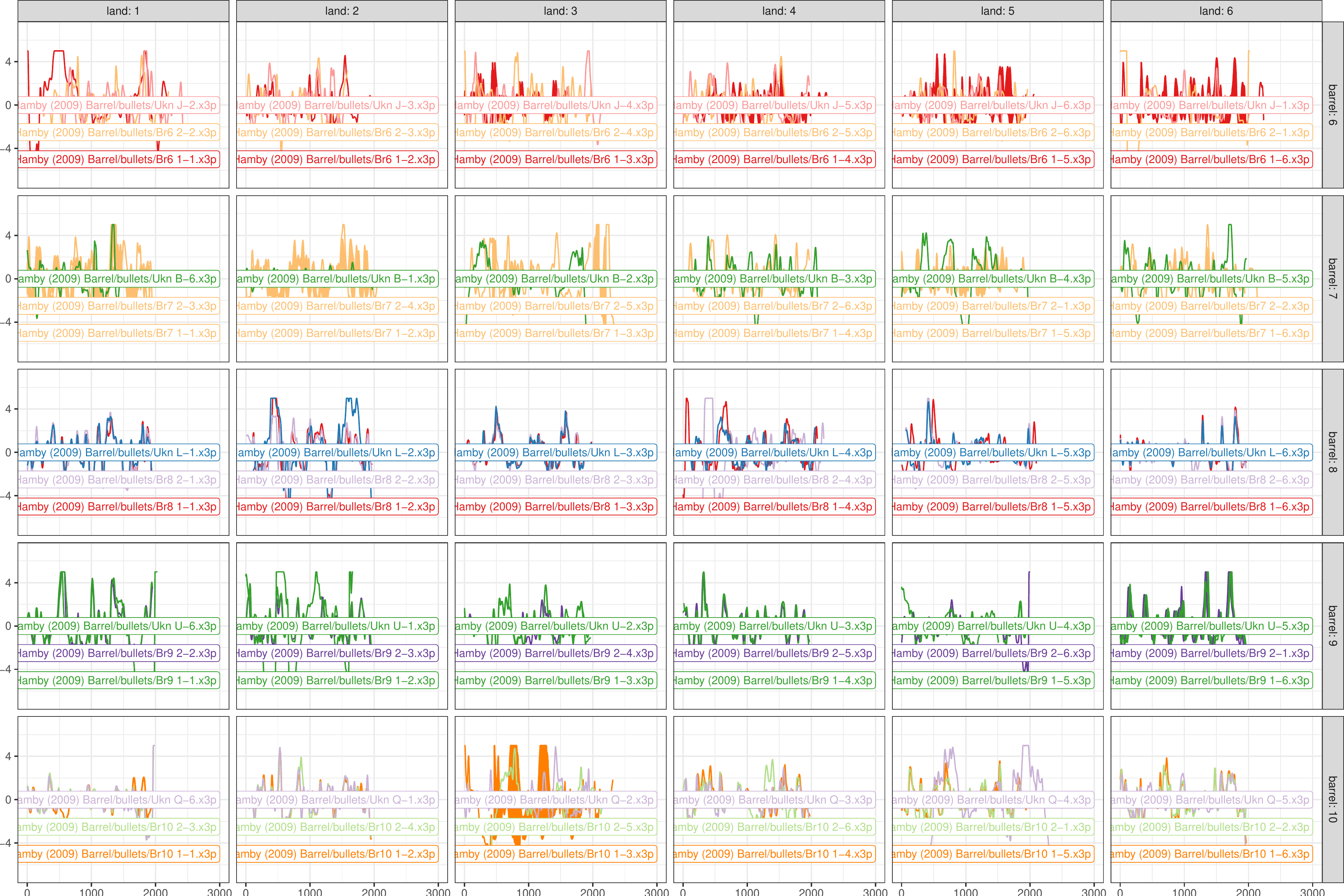}

\end{knitrout}
\caption{\label{fig:aligned-second}Overview of aligned signatures for all bullet land impressions for  barrels 5 to 10 of the Hamby study. }
\end{figure}
\end{landscape}
\newpage
\section{Table of feature importance}\label{supp:randomforest}
Two random forests were calculated for the Hamby study. For the first random forest only comparisons of bullet land impressions from known bullets and unknown bullets were used. The second random forest is based on a full comparison of every land impression with every other land impression, increasing the number of comparisons from originally 10,384 (10,212 known non-matches and 172 known matches) by another 10,931 comparisons (10,637 known non-matches and 94 known matches). Random forests allow an assessment of variable importance (also called feature importance) as the mean decrease in Gini index when including each variable.
Table~\ref{tab:importance} shows the results for feature importance for both of these random forests. Importance~1 refers to the smaller subset, Importance~2 is the feature importance derived from the random forest based on all pairwise comparisons.

\begin{table}[tbhp]
\caption{\label{tab:importance}Table of  features derived from bullet image ordered by importance in predicting matches. Importance is measured in terms of mean decrease in gini index when including the variable in a decision tree. Averages (and standard deviations) for known matches (KM) and known non-matches (KNM) are shown in the last four columns.}
\centering
\begin{tabular}{clrrrd{3.2}rd{3.2}}
  \hline
 & Variable & Importance 1 & Importance 2 & KM & (sd) & KNM & (sd).1 \\
  \hline
1 & ccf & 87.0 & 134.6 & 0.7 & ( 0.25) & 0.3 & (   0.10) \\
  2 & \#matches & 81.9 & 128.3 & 15.5 & ( 7.91) & 4.3 & (   2.51) \\
  3 & S & 46.4 & 53.6 & 18.3 & ( 8.95) & 5.5 & (   3.41) \\
  4 & \#non-matches & 35.9 & 62.7 & 9.8 & ( 5.80) & 18.8 & (   3.92) \\
  5 & D & 26.1 & 45.7 & 1.9 & ( 2.32) & 3.3 & (   1.94) \\
  6 & CMS & 15.9 & 25.1 & 8.5 & ( 5.65) & 2.3 & (   1.40) \\
  7 & CNMS & 13.3 & 20.9 & 4.8 & ( 4.10) & 10.2 & (   4.35) \\
   \hline
\end{tabular}

\end{table}

\bibliographystyle{imsart-nameyear.bst}
\bibliography{references}